\documentclass[%
 aip,
 amsmath,amssymb,
reprint,%
]{revtex4-2}
\usepackage{graphicx}
\usepackage{dcolumn}
\usepackage[T1]{fontenc}

\usepackage{bm}
\usepackage{cancel} 
\usepackage{subcaption}
\usepackage{hyperref}
\hypersetup{
    colorlinks=true,
    linkcolor=blue,
    filecolor=magenta, 
    citecolor=black,
    urlcolor=cyan,
    }

\usepackage{wrapfig}
\usepackage{ifpdf}
\usepackage{rotating}
\usepackage{alltt}
\usepackage{color}
\usepackage[usenames,dvipsnames]{xcolor}

\DeclareMathOperator*{\define}{\stackrel{def}{=}}

\newcommand{\eq}[1]{ Eq.\ (\ref{#1})}

\DeclareMathOperator*{\Stresstype}{\boldsymbol{P}}
\DeclareMathOperator*{\Heatfluxtype}{\boldsymbol{J}}

\DeclareMathOperator*{\StressIKone}{\boldsymbol{\Stresstype\limits^{\scriptscriptstyle{IK1}}}}
\DeclareMathOperator*{\StressMOP}{\boldsymbol{\Stresstype\limits^{\scriptscriptstyle{MoP}}}}
\DeclareMathOperator*{\StressCV}{\boldsymbol{\Stresstype\limits^{\scriptscriptstyle{CV}}}}
\DeclareMathOperator*{\PressureVA}{\boldsymbol{\Stresstype\limits^{\scriptscriptstyle{V\!A}}}}

\DeclareMathOperator*{\HeatfluxMOP}{{\Heatfluxtype\limits^{\scriptscriptstyle{MoP}}}}

\newcommand{\StressVIRIALinline}{%
  \mathop{\boldsymbol{\Stresstype}}%
  \nolimits^{\scriptscriptstyle VIRIAL}%
}

\newcommand{\StressIKoneinline}{%
  \mathop{\boldsymbol{\Stresstype}}%
  \nolimits^{\scriptscriptstyle IK1}%
}

\begin{document}

\preprint{AIP/123-QED}

\title{The Quantum Method of Planes - Local Pressure Definitions for Machine Learning Potentials}
\author{E. R. Smith}
 \email{edward.smith@brunel.ac.uk}
\affiliation{ 
Brunel University of London
Kingston Lane
Uxbridge
Middlesex UB8 3PH
}%

\date{\today}

\begin{abstract}
Stress, or pressure, is a central quantity in engineering and remains vital in molecular modelling. 
However, the commonly used virial stress tensor is invalid for an inhomogeneous fluid, which is essential in fluid dynamics and non-equilibrium molecular dynamics (NEMD) simulation. 
This is solved by using the method of planes (MoP), a mechanical form of pressure, simply interpreted as the force divided by area, yet is derived from the firm foundations of statistical mechanics. 
We present an extension of MoP stress \citep{Todd_et_al_95} to the MACE potential, a particular form of machine learning (ML) potentials allowing the incorporation of quantum mechanical (QM) physics into classical simulation. 
We present the derivation of this local stress for the MACE potential using the theoretical framework set out by \citet{Irving_Kirkwood}. 
For the test case of an interface between water and Zirconium Oxide, we show that the MoP measures the correct force balance while the virial form fails. 
Further, we demonstrate that this planar definition of stress is valid arbitrarily far from equilibrium, showing exact conservation every timestep in a control volume bounded by MoP planes. 
This links the stress directly to the conservation equations and demonstrates the validity in non equilibrium molecular dynamics (NEMD) systems. 
All code to reproduce these validations for any MACE system, together with ASE accelerated code to calculate the MoP, is provided as open source. 
This work helps build the foundation to extend the ML revolution in materials to NEMD and molecular fluid dynamics modelling.
\end{abstract}

\maketitle

\section{Introduction}

Material science is undergoing a revolution, with a new generation of machine learning (ML) potentials allowing classical models to incorporate quantum Density Functional Theory (DFT) level accuracy at speeds traditionally closer to classical molecular dynamics \citep{Eyert2023}. One promising candidate for fast and accurate simulation is the MACE-MP-0 family of models \citep{batatia2023foundation}, which are trained on an extensive material database covering the periodic table and giving good results for cases outside of this dataset, including the mixing of elements. 
The MACE model is particularly efficient and promising because, rather than a purely data driven approach that simply feeds vast amounts of data to an increasing deep neural network, it uses aspects of the physics to simplify the required network and training \citep{Kovacs_et_al2023}.
This speeds up inference, the force calculation in MACE MD, to speeds approaching those of classical MD when run on modern GPU architectures.
The symmetries of common molecular configurations during interaction are built in through the atomic cluster expansion (ACE) \citep{RDrautz_2004}, which uses symmetries to reduce the number of different molecules and configurations that need to be remembered by a network. 
This can be understood as analogous to machine vision, where recognising a picture of a cat, even when shifted or rotated, is still the same picture.
In molecules, this can be done using symmetries through the E3NN library \citep{weiler20183dsteerablecnns, kondor2018clebschgordannets, thomas2018tensorfieldnetworks, e3nn_software}.
The use of ACE allows high body order \citep{RDrautz_2004}, the number of many-body interactions considered (in MACE set to a bond order of four locally).  
Inspired by leading results in message passing potentials like NequIP \citep{batzner2022e3gnn}, this ACE approach is then embedded in a message passing framework. 
The required cutoff length of the ACE molecular interactions, $r_{c}$, is reduced using this message passing approach (the M in MACE), which splits multi-body interactions into a graph network so only energetically important connections or "edges" are required and many-body interactions occur indirectly through hops of the graph network.
The number of hops taken is $M=2$ on the MACE-MP-0 version used in this work.  \footnote{The $1$ and $3$ were found to be less favourable in terms of accuracy vs. computational cost. As MACE uses a more complicated local potential with ACE, it requires fewer hops than other message passing approaches.}.
A machine learning process is then applied to work out which connections are essential, a process which means each four-body local calculation is connected by graph edges to $M$ other atoms, each with four body interactions given by the ACE model. 
The resulting potential is then effectively $13$ body order with an effective cutoff length $M r_{c}$. 
As a result, this performs very well in learning DFT results while allowing fast simulations.
With the simplifications of M and ACE, a deep neural network is then trained on the MPtraj \citep{deng_2023_chgnet} and sAlex \citep{barroso_omat24} databases, covering many materials.
\citet{batatia2023foundation} demonstrate that the MACE model gives good results on a vast range of cases. 
This has since been extended to include other databases, charges, fine tunes and architecture tweaks since initial publications \citep{batatia2025}, a process which will doubtlessly accelerate.
Such fine-tuning also has the ability to improve properties DFT typically struggles to reproduce, from dispersion, long-range interactions to the over structuring of water.
However, the presented work is not tied to the specific weights of this foundation model, which has already been fine-tuned and improved since initial publication using wider DFT databases.
Instead, the derivation in this work focuses on the base architecture so should work with the most up to date model with similar architecture.

As a result of the potential revolutionary nature of ML potentials, and the already significant improvement offered by the MACE in materials modelling, the solid simulation community has fully embraced these models (see, for example, the range of authors on \citet{batatia2023foundation}). 
However, the potential of ML for fluid dynamics problems, known as non equilibrium molecular dynamics (NEMD), seems to have received far less attention.
Modelling of the interface between a liquid and a surface is often a very complex problem that would greatly benefit from ML modelling.
Behaviour such as rusting, catalysis or water dissociation are difficult to capture with classical models like the ReaxFF reactive force-field \citep{Senftle2016}. 
Cooling of heat exchanges, nuclear reactors or semiconductors all require molecular interface modelling, while lubrication and tribology are intertwined with the molecular chemistry of the interface.
Using ML models to capture these effects has the potential to revolutionise NEMD modelling of interfaces.
NEMD is characterised by a series of techniques to study fluid flows, including thermostats to control temperature or forcing routines to enforce shear flow \citep{todd_daivis_2017}.
NEMD is molecular fluid dynamics and so provides methods to get quantities of great importance to fluid flow modelling, including density, velocity, stress and heat flux \citep{Evans_Morris, todd_daivis_2017}.
Of particular importance for many cases in fluid dynamics are the stress tensor and heat flux vectors, vital for rheology and tribology \citep{Ewen2017} with active research in slip length \citep{Kannam2012, Williams2022} and the moving three-phase contact line \citep{Thompson1989, Qian2005}.
Pressure is also essential for surface tension \citep{Kirkwood_Buff}, viscosity and heat flux \citep{Green,Kubo}.
These molecular details can be included in continuum-based engineering simulation, such as computational fluid dynamics (CFD), through coupled simulation \citep{OConnell1995, Mohamed2010} where stress coupling is useful in both fluid simulation \citep{Flekkoy2000} and for solid mechanics \citep{Davydov2014}.

Getting the stress and heat flux has a long history in the literature \citep{Shi_et_al(2023)}.
Two key observations are important: stress is non-unique \citep{Schofield_Henderson,Admal_Tadmore_2010,Heinz_et_al05} and certain forms of stress are incorrect for inhomogeneous systems \citep{todd_daivis_2017, heyes2011equivalence}.
Note that stress and pressure are used interchangeably in this work, with one simply the negative of the other.
For the NEMD cases, a stress and heat flux that is valid away from thermodynamic equilibrium is essential.
Recent work by \citet{langer2023stress}, \citep{Langer2023heatflux}, demonstrates that stress is possible to obtain for a whole periodic system by applying a finite strain.
The current work extends this by starting from the foundations of \citet{Irving_Kirkwood} to derive a local form of stress and heat flux but expressed here in integrated or weak control volumes form \citep{Smith_et_al12}.
This results in the Method of Planes MoP stress \citep{Todd_et_al_95} which can be shown to be linked exactly to a control volume form of the conservation equations.
This link to a conservative form provides the validation that the stress defined is meaningful and correct.
In this process, we also show that the virial form of stress and heat flux, the default option in many MD packages, are an oversimplification that lead to errors away from equilibrium.
This is a well-know result in the classical literature \citep{todd_daivis_2017}, but extended here to ML potentials.
The MoP also has the advantage it is the most fundamental definition of stress, the force divided by area, and can be calculated relatively easily in MD.
As a result, this work will derive a usable stress for the MACE potential and demonstrate the conservation of momentum and energy in a control volume. 
More generally, this treatment will work for any machine learning potential that can be expressed as a pairwise force, which includes the general class of message passing neural networks \citep{langer2023stress}.

The remainder of the work is as follows, in section \ref{sec:theory} a derivation of the MoP stress and heat flux for an ACE style potential is presented. Next, the methodology of the molecular dynamics used to simulate a nontrivial case of ZrO$_2$ and water is given in section \ref{sec:methodology}. The results of applying the MoP stress in this system is shown in section \ref{sec:results} including a demonstration of the conservation of momentum in a control volume and the failure of the virial form of stress. Finally, some brief conclusions are given in section \ref{sec:conclusions}.

\section{Theory}
\label{sec:theory}

In classical physics, the force $\boldsymbol{F} = - \boldsymbol{\nabla} U(\boldsymbol{r})$, is given by the negative gradient of the potential $U$ with respect to position $\boldsymbol{r}$ only if we have a conservative potential \citep{Goldstein}.
In classical molecular dynamics we have a system of pointwise atoms so only at the positions of these atoms $ \boldsymbol{r}_i $ can forces exist.
Then the force on atom $i$ is,
\begin{align}
\boldsymbol{F}_i = -\frac{\partial U}{\partial \boldsymbol{r}_i}
\end{align}
with the total forces in an MD system satisfying the condition $\sum_{i=1}^N \boldsymbol{F}_i = 0$.
The true picture in a quantum system is more complex with wavefunctions defining a continuous energy throughout space, which is reduced by ML to a graph network.
It is interesting to note that the existence of a conservative potential, while generally satisfied by just training for energy only, is better enforced by the explicit inclusion of forces in the training process.
The training process of MACE aims to match both total potential energy $U$ and forces $\boldsymbol{F}_i$ to DFT data \citep{Kovacs_et_al2023}, by minimising the function,
\begin{align}
\mathcal{L} = \frac{\lambda_E}{B} \sum_{b=1}^{B} \left[\frac{U_b - \widetilde{U}_b }{N_b} \right]^2 + \frac{\lambda_F}{3B} \sum_{b=1}^{B}  \sum_{i=1}^{N_b} \left[-\frac{\partial U_b}{\partial \boldsymbol{r}_i} - \widetilde{\boldsymbol{F}}_{ib} \right]^2 
\end{align}
where the index $b$ denotes the sum over the $B$ total batches used in the training process (here originally for MACE-MP-0 the 89 elements in the MATPES-PBE dataset \citep{kaplan2025foundational}). The $\widetilde{U}_b$ and $\widetilde{\boldsymbol{F}}_{ib}$ are the potential energies and forces from batch $b$ of the DFT data respectively.
The Lagrangian multipliers $\lambda_E$ and $\lambda_F$ are chosen to enforce the relative importance of the two contributions, where it was found an initial training using $\lambda_E >> \lambda_F$ before switching to $\lambda_E << \lambda_F$ with a smaller step gave good agreement for energy and forces \citep{Kovacs_et_al2023}.

\subsection{Intermolecular Forces}

\citet{Admal_Tadmor_2011} argue all many-body forces can be written as "forces parallel to the directions connecting the particle to its neighbours", provided the potentials are \textit{continuous}.
For classical potentials, for example the Lennard-Jones potential, the continuity is true due to the well defined mathematical function of the potential, i.e. $U(r) = r^{-12} - r^{-6}$.
The MACE model follows the general framework of message passing neural networks (MPNNs) \citep{batatia2022mace}, a type of graph-based machine-learning potential.
As a result, these MPNN are \textit{continuous} and can use automatic differentiation (AD) to define gradients, so can be expressed in terms of pairwise forces \citep{langer2023stress}.
The message passing uses ACE, which is also constructed from a set of \textit{continuous} functions expressed in a pairwise manner. 
For machine learning potentials, in particular the ACE potential used here,
the internal energy is typically defined using set notation \citep{batatia2022mace, langer2023stress, Kovacs_et_al2023} $U = U\left( \{ \boldsymbol{r}_{ij} \; | \; r_{ij} < r_c \} \right)$ to denote the potential energy is a function of all possible pairwise interactions between all atoms within cutoff $r_c$.
With a cutoff, the sum is only over the atoms within the limit of this cutoff, often denoted by set notation in the ML literature \citep{langer2023stress}, e.g. $\mathcal{N}(i) = \{j \ne i | r_{ij} < r_c \}$ for all atomic "senders" to atom $i$.
For more complex geometric graphs or message passing cases, it might occur that interactions $\mathcal{N}(i)$ includes a given $j$ where $i$ is not in the set of $j$'s interactions $\mathcal{N}(j)$ ($j$ is not a receiver from $i$).    
For the cases used in this work, all receivers and senders are balanced with only pairwise interactions within the cutoff included and many-body contributions are handled by the hidden variables.
This means the sums in \eq{Forces_pairwise} can simply be written over all $N$ interactions provided $j \ne i$ without using set notation.
As a result,
taking the derivative of this potential proceeds as follows,  
\begin{align}
\frac{\partial U}{\partial \boldsymbol{r}_i} &  =  \sum_{j = 1 }^N \sum_{k \ne j }^N 
\frac{\partial U}{\partial \boldsymbol{r}_{jk}} \cdot \frac{\partial \boldsymbol{r}_{jk}}{\partial \boldsymbol{r}_i} 
\nonumber \\
&  = \underbrace{\sum_{j \ne i }^N \frac{\partial U}{\partial \boldsymbol{r}_{ji}} \cdot \frac{\partial \boldsymbol{r}_{ji}}{\partial \boldsymbol{r}_i}}_{\text{Case where } k=i}  + \underbrace{\sum_{k \ne i }^N \frac{\partial U}{\partial \boldsymbol{r}_{ik}} \cdot \frac{\partial \boldsymbol{r}_{ik}}{\partial \boldsymbol{r}_i}}_{\text{Case where } j=i}
\nonumber \\
&  = \sum_{j \ne i }^N \left[
\frac{\partial U}{\partial \boldsymbol{r}_{ij}} \cdot \frac{\partial \boldsymbol{r}_{ij}}{\partial \boldsymbol{r}_i} 
+ \frac{\partial U}{\partial \boldsymbol{r}_{ji}} \cdot \frac{\partial \boldsymbol{r}_{ji}}{\partial \boldsymbol{r}_i}  \right]
\nonumber \\
& = \sum_{j \ne i }^N \left( 
\frac{\partial U}{\partial \boldsymbol{r}_{ij}} - \frac{\partial U}{\partial \boldsymbol{r}_{ji}} 
\right) ,
\label{Forces_pairwise}
\end{align}
Here the derivative of the potential is expanded to every possible pairwise interaction distance, although in practice for MACE this would only be interactions on the graph network.
Only when $j=i$ or $k=i$ are the derivatives non-zero and then we relabel dummy index $k$ to get the third line.
Finally, as $ \boldsymbol{r}_{ij} = \boldsymbol{r}_i - \boldsymbol{r}_j $, the derivatives w.r.t to $\boldsymbol{r_i}$ equal $1$ ($-1$ for $\boldsymbol{r}_{ji})$.
We can then define the final expression in \eq{Forces_pairwise} as the antisymmetric pairwise force $\boldsymbol{f}_{ij}$,
\begin{equation}
\boldsymbol{f}_{ij} \define \frac{\partial U}{\partial \boldsymbol{r}_{ij}} - \frac{\partial U}{\partial \boldsymbol{r}_{ji}}
\label{Enforce_Newton3rd}
\end{equation}
The total force on atom $ i $ is then simply
 $\mathbf{F}_i = \sum_{j \ne i} \boldsymbol{f}_{ij}$ 
Even though MACE is a non-pairwise (many-body) potential, with energy depending on complex correlations, the forces can be decomposed into pairwise sums that match the graph network.
The chain rule operates on the input vectors in \eq{Forces_pairwise} and regardless of how complex the function $U\left( \{ \boldsymbol{r}_{ij} \} \right)$ inside the neural network, the gradient is eventually distributed back to the edge vectors of the graph within the cutoff length $r_c$.
However, it is still true that $\partial U/ \partial \boldsymbol{r}_{ij} \ne - \partial U/ \partial \boldsymbol{r}_{ji}$ in general.
For the many-body potential modelled by MACE, the amount of energy the system changes by if $i$ moves closer to $j$ ($\partial U / \partial \boldsymbol{r}_{ij}$) is not the same as the amount of energy the system changes by if $j$ moves closer to $i$ ($\partial U / \partial \boldsymbol{r}_{ij}$).
These two calculations depend on different parts of the graph network.
\citet{Admal_Tadmore_2010} state this difference succinctly; noting pairwise forces can be physically interpreted as the ``force exerted on particle $i$ by particle $j$'', whereas the more general interaction force $\boldsymbol{f}_{ij}$ can be interpreted as the ``contribution to the force on particle $i$ due to the presence of particle $j$''.
However, the resulting form of \eq{Enforce_Newton3rd} guarantees $ \boldsymbol{f}_{ij} = -\boldsymbol{f}_{ji} $ and so satisfies Newton's third law.
In a graph neural network, molecules are at the nodes while the connections between them, called the graph edges, conceptually correspond to the intermolecular interactions.
Note that in the work of \citet{langer2023stress} they explicitly denote that $\boldsymbol{r}_{ij}$ applies the minimum image convention which is dropped here for notational simplicity. 

It is instructive to consider the pseudo code used to get $\boldsymbol{f}_{ij}$ of \eq{fij_def} from a graph neural network by autodifferentiation, which proceeds as follows.


\begin{verbatim}
#Get arrays with all interacting atoms stored 
#in identical sized arrays of corresponding 
#sender/reciever indices
sender, receiver = neighbour_list(r, cutoff)

#Positions of ri and rj at the end of every 
#intermolecular interaction or graph "edge" 
#give the rij vectors (optionally add a shift 
#for periodic boundaries)
rij = r[receiver] - r[sender]

#Taking gradient w.r.t vector rij
dUdrij = autograd(energy, rij)

#Reshape n_edges to n_nodes^2 
fij[sender, receiver, :] = -dUdrij

#Apply anti-symmetry operation
fij[:,:,0] = fij[:,:,0] - fij[:,:,0].T
fij[:,:,1] = fij[:,:,1] - fij[:,:,1].T
fij[:,:,2] = fij[:,:,2] - fij[:,:,2].T
\end{verbatim}

The process of putting all pairwise forces into an N by N matrix reduces computational efficiency gains provided by the MPNNs architecture, which is designed to only include relevant interactions.
To improve this approach, a sparse matrix could be used for fij or the algorithm be used with only the sender and receiver MPNNs interactions.
However, the above code is reasonable for small systems given the force calculation is often the major component in simulation cost and obtaining stresses is undertaken only as an intermittent exercise.
Generally in NEMD we allow the system to evolve sufficiently that the measured stress is statistically independent of previous values.
The stress autocorrelation time of the system is usually a sensible value for the sample frequency as it ensures stress measurements are statistically uncorrelated.
In section \ref{sec:MomEqn} we use the form of inter-molecular force in \eq{Enforce_Newton3rd} to determine the pressure tensor

It is worth making a comment on the link to the central force decomposition of \citet{Admal_Tadmore_2010}.
Any choice of pairwise force is not unique and MACE uses a different choice than the central force decomposition.
MACE does not satisfy the assumption of a central force decomposition \citep{Admal_Tadmore_2010} that $\partial U/\partial \boldsymbol{r}_{ij} = \partial U/\partial r_{ij} \widetilde{\boldsymbol{r}}_{ij}$ with $\widetilde{\boldsymbol{r}}_{ij}$ the unit vector between $i$ and $j$.
This can be quickly checked by dotting the MACE forces with the unit vector $\widetilde{\boldsymbol{r}}_{ij}$, showing $\partial U/\partial \boldsymbol{r}_{ij} \cdot \widetilde{\boldsymbol{r}}_{ij} \ne \partial U/\partial r_{ij} $.
\citet{admal2014results} discusses in his thesis that pairwise classical mechanical $\boldsymbol{f}_{ij}$ forces are at odds with Quantum mechanics where only $\boldsymbol{F}_{i}$ and $\boldsymbol{F}_{j}$ can be meaningfully obtained.
This can be taken together with the \citet{Admal_Tadmore_2010} demonstration that with 5 or greater bodies the decomposition is no longer unique.
The MACE potential used in this work is based on an effective body order of 13 with message passed variables and graph hops to retain many-body DFT detail encoded in $\boldsymbol{f}_{ij}$.
The pairwise force presented is therefore greater than five body and so cannot be mathematically unique. 
\citet{Admal_Tadmore_2010} show this condition that forces act along $\widetilde{\boldsymbol{r}}_{ij}$ is required for conservation of angular momentum.
Angular momentum conservation in MACE is said to be enforced by the symmetrical construction of the E3NN architecture \citep{Brandstetter2021GeometricAP}, and this conservation has been checked in recent work \citep{Langer_2024}.
Angular and linear momentum conservation is enforced for the whole system by this E3NN architecture, but not for any given pairwise interaction.
Linear momentum conservation for pairwise interactions are enforced by Newton's 3rd law, which is not required but the definition of \eq{Enforce_Newton3rd} specifically enforces this symmetry.

The definition of pairwise force in \eq{Enforce_Newton3rd} follows directly from the assumption $U$ is a function of pairwise forces only. 
\citet{fan2015force} show the three body potentials including Tersoff, Brenner, Stillinger-Weber and a general many-body potential as simply $\boldsymbol{f}_{ij} = \partial U_i / \partial \boldsymbol{r}_{ij} - \partial U_j / \partial \boldsymbol{r}_{ji}$ where $U_i$ is the energy at the point location of particle $i$. 
The assumption of individual atomic energy $U = \sum_{i=1}^N U_i$ in $\partial U/\partial \boldsymbol{r}_i$ can also be used to define an alternative version of pairwise forces as,
\begin{align}
\tilde{\boldsymbol{f}}_{ij} \define - \frac{\partial U_j}{\partial \boldsymbol{r}_i }
\label{fij_def}
\end{align}
This version of force is shown to be the required form to conserve energy by \citet{Langer2023heatflux}, as derived here in section \ref{sec:energy} and demonstrated to work locally in section \ref{sec:results}, although it will not satisfy Newton's 3rd law in general as $\partial U_j/\partial \boldsymbol{r}_i \ne \partial U_i/\partial \boldsymbol{r}_j$ .

.

\subsection{Momentum Equation}
\label{sec:MomEqn}

We use the definition of forces from \eq{Enforce_Newton3rd} in the derivation of pressure following the statistical mechanical process of \citet{Irving_Kirkwood}.
Assuming phase space is bounded, \citet{Irving_Kirkwood} obtain an expression for the evolution in time of expected value $\alpha$,  
 \begin{align}
\frac{\partial }{\partial t} \bigg\langle \alpha ; \textit{f} \bigg\rangle  = \displaystyle\sum_{i = 1 }^{N} \bigg\langle \boldsymbol{F}_i \cdot \frac{\partial \alpha}{\partial \boldsymbol{p}_i} +\frac{\boldsymbol{p}_i}{m_i} \cdot \frac{\partial \alpha}{\partial \boldsymbol{r}_i} ;\textit{f} \bigg\rangle,
\label{Eq_IKalpha_evo}
\end{align}
where $\textit{f}$ is the probability of finding a molecule at a point in phase space and the inner product with $\langle\alpha; f\rangle$ represents the ensemble average of $\alpha$.
By letting $\alpha = \sum \boldsymbol{p}_i  \delta(\boldsymbol{r}_i-\boldsymbol{r})$, \citet{Irving_Kirkwood} define the momentum density at a point in space,
\begin{align}
\rho \boldsymbol{u} \define \displaystyle\sum_{i=1}^{N} \bigg\langle \boldsymbol{p}_i  \delta(\boldsymbol{r}_i-\boldsymbol{r}) ; \textit{f} \bigg\rangle,
\label{Momdensity}
\end{align}
The time evolution of momentum is used to obtain the pressure tensor by taking the time derivative of both sides of \eq{Momdensity} and applying \eq{Eq_IKalpha_evo},
\begin{align}
\frac{\partial }{\partial t}  \rho \boldsymbol{u} & = \frac{\partial }{\partial t}  \displaystyle\sum_{i=1}^{N}  \bigg\langle \boldsymbol{p}_i \delta(\boldsymbol{r}_i-\boldsymbol{r}) ; \textit{f} \bigg\rangle \nonumber \\
& = \displaystyle\sum_{i = 1 }^{N} \bigg\langle \boldsymbol{F}_i \delta(\boldsymbol{r}_i-\boldsymbol{r}) - \frac{\partial }{\partial \boldsymbol{r}} \cdot \frac{\overline{\boldsymbol{p}}_i \overline{\boldsymbol{p}}_i }{m_i}  \cdot   \delta(\boldsymbol{r}_i-\boldsymbol{r}) ;\textit{f} \bigg\rangle 
\label{IK_timeevo}
 \end{align}
The second term is the kinetic part of the pressure tensor $\boldsymbol{P}^k$, defined using $\partial / \partial \boldsymbol{r}_i  \delta(\boldsymbol{r} - \boldsymbol{r}_i) =- \partial/ \partial \boldsymbol{r} \delta(\boldsymbol{r} - \boldsymbol{r}_i) $ and subtracting the streaming velocity.
As $\boldsymbol{p}_i$ is the momentum in the laboratory frame we can denote $\overline{\boldsymbol{p}}_i$ as the peculiar value which excludes the macroscopic streaming term $\boldsymbol{u}(\boldsymbol{r}_i)$ at the location of molecule $i$ \citep{Evans_Morris}. 
The kinetic pressure integrated over a volume is $\int_V \boldsymbol{P}^k dV = \langle \frac{\overline{\boldsymbol{p}}_i \overline{\boldsymbol{p}}_i }{m_i}  \vartheta_i ;\textit{f} \rangle$ where the function $\vartheta_i$ is the integral of a Dirac delta function.
This is a function which has a value of one when a molecule is inside the volume and zero otherwise.
The $\vartheta_i$ has a clearly defined mathematical form from the integral of the Dirac delta function between the finite limits of a volume, for example a cuboid between ($\alpha^+$ and $\alpha^-$ for $\alpha \in \{x,y,z \}$) then 
$\vartheta_i = \prod_{\alpha \in \{x,y,z \}} [H(\alpha^+ - \alpha_i) -  H(\alpha^- - \alpha_i)]$, where $\prod$ denotes the product operator taken here over the 3 spatial dimensions.
This is an expression which selects molecules inside the Heaviside functions.
The kinetic term is identical in both classical and MACE systems so is not considered further, instead we turn our attention to the configurational pressure.
We aim to express the forcing term of \eq{IK_timeevo} as the divergence of a pressure tensor $\boldsymbol{\nabla} \cdot \boldsymbol{P}^c$. 
To do this, \citet{Irving_Kirkwood} use the assumption of Newton's 3rd law, which we can also invoke due to the definition of \eq{Enforce_Newton3rd}, to get the difference of two Dirac delta functions,
\begin{align}
\displaystyle\sum_{i = 1 }^{N} & \bigg\langle \boldsymbol{F}_i \delta(\boldsymbol{r}_i-\boldsymbol{r}) ;f \bigg \rangle  = -\displaystyle\sum_{i = 1 }^{N} \bigg\langle \frac{\partial U}{\partial \boldsymbol{r}_i} \delta(\boldsymbol{r}_i-\boldsymbol{r}) ;f \bigg \rangle    \nonumber \\
&  = -\cfrac{1}{2}\bigg\langle \displaystyle\sum_{i = 1 }^{N} \frac{\partial U}{\partial \boldsymbol{r}_i} \delta(\boldsymbol{r}_i-\boldsymbol{r})  + \displaystyle\sum_{j = 1 }^{N}  \frac{\partial U}{\partial \boldsymbol{r}_j} \delta(\boldsymbol{r}_j-\boldsymbol{r});f \bigg \rangle \nonumber \\
& = -\cfrac{1}{2}\bigg\langle \displaystyle\sum_{i = 1 }^{N}  \sum_{j \ne i}^N \left( 
\frac{\partial U}{\partial \boldsymbol{r}_{ij}} - \frac{\partial U}{\partial \boldsymbol{r}_{ji}} 
\right)  \delta(\boldsymbol{r}_i-\boldsymbol{r})  \nonumber \\ 
& \;\;\;\;\;\;\;\;\;\;\;\;\;\;+ \displaystyle\sum_{j = 1 }^{N}  \sum_{i \ne j}^N \left( 
\frac{\partial U}{\partial \boldsymbol{r}_{ji}} - \frac{\partial U}{\partial \boldsymbol{r}_{ij}} 
\right)  \delta(\boldsymbol{r}_j-\boldsymbol{r});f \bigg \rangle  \nonumber \\
& = \cfrac{1}{2}\displaystyle\sum_{i=1}^{N}  \displaystyle\sum_{j \ne i}^{N} \bigg\langle   \boldsymbol{f}_{ij} \delta(\boldsymbol{r}_i-\textbf{r}) + \boldsymbol{f}_{ji} \delta(\boldsymbol{r}_j-\boldsymbol{r}) ;f \bigg \rangle \nonumber \\
& = \cfrac{1}{2}\displaystyle\sum_{i=1}^{N}  \displaystyle\sum_{j \ne i}^{N} \bigg\langle   \boldsymbol{f}_{ij} \left[\delta(\boldsymbol{r}_i-\textbf{r}) - \delta(\boldsymbol{r}_j-\boldsymbol{r})\right] ;f \bigg \rangle
\label{force_term}
\end{align}
At this point, a useful insight can be obtained from the difference between the two volume-integrated Dirac Delta functions, $\vartheta_{ij} = \vartheta_i - \vartheta_j$.
Equation \eq{force_term} integrated over a volume is,
\begin{align}
\displaystyle\sum_{i = 1 }^{N} & \bigg\langle \boldsymbol{F}_i\vartheta_i ;f \bigg \rangle  = \cfrac{1}{2}\displaystyle\sum_{i=1}^{N}  \displaystyle\sum_{j \ne i}^{N} \bigg\langle   \boldsymbol{f}_{ij} \vartheta_{ij} ;f \bigg \rangle
\label{varthetaij}
\end{align}
The term $\vartheta_{ij}$ is only non zero when molecule $i$ is in the volume ($\vartheta_i=1$) and molecule $j$ is not ($\vartheta_j=0$), or vice versa giving a negative sign.
With both inside, $\vartheta_{ij}  = 1 - 1 = 0$ or outside $\vartheta_{ij} = 0 - 0$.
Physically this can be understood for molecule $i$ as only the interaction with molecules outside of its current volume being able to change the momentum or energy that $i$ adds to that volume.
All interactions inside the volume are equal and opposite so result in no change in the volume's momentum or energy.
As the volume is a closed surface, by construction, these $\vartheta_{ij}$ interactions must be surface crossings.
In this form, the balance on an arbitrary volume can be used to check momentum conservation where surface forces equal internal momentum change (given the absence of any atoms crossing the surface).
Stress is a more useful form, so the usual manipulations can be made, including the slightly tenuous Taylor expansion of two Dirac delta functionals from \citet{Irving_Kirkwood}, to give $O_{ij}$ the so-called IK operator as an expansion in delta functions \citep{Todd_et_al_95}, or the more useful form obtained by rewriting the integral along a (non unique \citep{Schofield_Henderson}) contour \citep{Noll1955, Monaghan1997},
\begin{align}
\delta(\boldsymbol{r}_i-\textbf{r}) - \delta(\boldsymbol{r}_j-\boldsymbol{r}) 
& =  - \boldsymbol{r}_{ij} \cdot \frac{\partial}{\partial \boldsymbol{r}} O_{ij} \delta(\boldsymbol{r}_i-\boldsymbol{r})
\nonumber \\
& = \oint \frac{\partial}{\partial \boldsymbol{\ell}}  \cdot \delta \left({\boldsymbol{r}} - \boldsymbol{r}_i - \boldsymbol{\ell} \right)  d\boldsymbol{\ell} 
\nonumber \\
& = \frac{\partial}{\partial {\boldsymbol{r}}} \cdot  \oint  \delta \left({\boldsymbol{r}} - \boldsymbol{r}_i - \boldsymbol{\ell} \right)  d\boldsymbol{\ell} .
\label{SHContour}
\end{align}
We can go further and simply assume a straight line, often called the IK contour which is consistent with Newton’s definition of what he calls the "impressed force" between two points,
\begin{align}
\oint  \delta \left({\boldsymbol{r}} - \boldsymbol{r}_i - \boldsymbol{\ell} \right)  d\boldsymbol{\ell} 
\approx  \boldsymbol{r}_{ij} \int_{0}^{1}  \delta \left({\boldsymbol{r}} - \boldsymbol{r}_i - \lambda \boldsymbol{ r_{ij}} \right)   d\lambda.
\label{LinContour}
\end{align}
This assumes $\boldsymbol{\ell} = \lambda \boldsymbol{ r_{ij}}$ with $0 < \lambda < 1$ so 
$d\boldsymbol{\ell} = \boldsymbol{ r_{ij}} d\lambda $.
MACE is trained on the underlying DFT models of the full interaction of electronic wavefunctions, which in classical mechanics we simplify to interacting potentials between point atoms.
Although the form of MACE is a graph network of linear atom connection, the actual force obtained from the sum of these connections represents this more complex DFT calculation from a many-body environment.
As a result, the assumption of pairwise linear interactions between atoms is oversimplified \footnote{ }, and we instead keep the unspecified contour form from \eq{SHContour} and substitute into \eq{force_term} .
This is then integrated over a finite volume to get configurational pressure,
\begin{align}
\frac{\partial}{\partial {\boldsymbol{r}}}  & \int_V \cdot \boldsymbol{P}^C dV \nonumber \\
& =  \frac{\partial}{\partial {\boldsymbol{r}}} \cdot \frac{1}{2}   \displaystyle\sum_{i=1}^{N} \displaystyle\sum_{j \ne i}^{N}  \int_V \bigg\langle   \boldsymbol{f}_{ij}  \oint  \delta \left({\boldsymbol{r}} - \boldsymbol{r}_i - \boldsymbol{\ell} \right)  d\boldsymbol{\ell} ; f \bigg\rangle  dV 
\nonumber \\
& =  \frac{\partial}{\partial {\boldsymbol{r}}} \cdot \frac{1}{2}    \displaystyle\sum_{i=1}^{N} \displaystyle\sum_{j \ne i}^{N}  \bigg\langle   \boldsymbol{f}_{ij}  \oint \vartheta_{\ell}  d\boldsymbol{\ell}  ; f \bigg\rangle 
\label{Config_pressure}
\end{align}
the function $\vartheta_{\ell }$ is non zero if the part of the non-linear interaction path is inside the averaging volume \citep{Smith_et_al12}.
The full volume averaged pressure can be obtained by assuming constant pressure in the volume,
\begin{align}
	\PressureVA = \frac{1}{\Delta V}  \bigg\langle  \displaystyle\sum_{i = 1 }^{N} \frac{\overline{\boldsymbol{p}}_i \overline{\boldsymbol{p}}_i}{m_i}  \vartheta_i + \frac{1}{2}\displaystyle\sum_{i=1}^{N} \displaystyle\sum_{j \ne i}^{N}    \boldsymbol{f}_{ij} \boldsymbol{l}_{ij} ; f \bigg\rangle,
	\label{VA_stress}
\end{align}
where $\Delta V$ is the local volume and $\boldsymbol{l}_{ij} =  \oint \vartheta_{\ell}  d\boldsymbol{\ell}$ is a vector integral along the interaction taking the fraction inside a volume.
This form is well know in the literature for straight lines \citep{Cormier_et_al, Hardy}, and has been written here for a contour. 

\begin{figure*}
\includegraphics[width=0.95\textwidth]{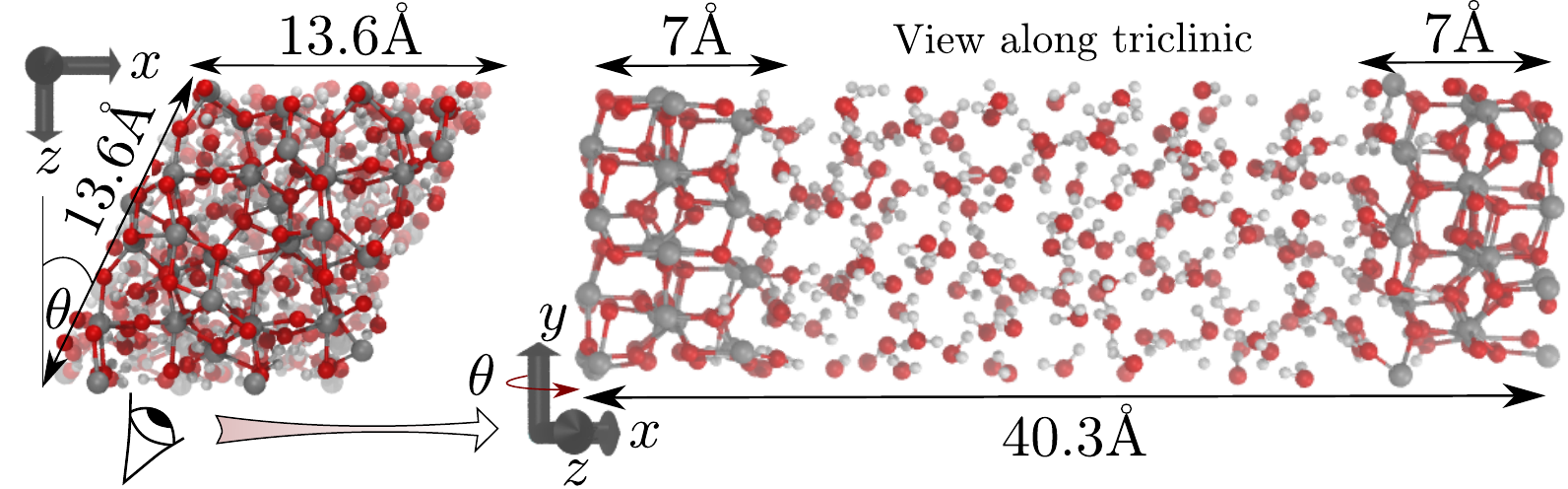}
\caption{The domain with Zirconium dioxide(ZrO$_2$) walls and water (H$_2$O) in the channel showing all dimensions. The top is shown on the left highlighting the triclinic nature of the system, with the view angle on the right along the tricilinc domain angle of about $\theta = 9^{\circ}$ to highlight the crystal structure. The domain is directly taken from the DFT work of \citet{Yang_et_al2021} with the fluid region doubled in size by copying the molecules and re-equilibrating the larger system.}  
\label{fig:domain}
\end{figure*}

The virial form of pressure \citep{Parker} is a simplification of \eq{VA_stress} where instead of assigning to a volume based on the path of interactions, the pressure contribution is split with half assigned to each atom. 
Applying this pairwise decomposition to a local volume gives,
\begin{align}
	\StressIKone = \frac{1}{\Delta V} \displaystyle\sum_{i = 1 }^{N} \bigg\langle   \left( \frac{\overline{\boldsymbol{p}}_i \overline{\boldsymbol{p}}_i}{m_i}  + \frac{1}{2}\displaystyle\sum_{j \ne i}^{N}    \boldsymbol{f}_{ij} \boldsymbol{r}_{ij} \right)\vartheta_i ; f \bigg\rangle.
	\label{virial}
\end{align}
The term in \eq{virial} is called the IK1 pressure, following notation in \citet{Evans_Morris}, as it represents the truncation of the \citet{Irving_Kirkwood} expression at the first term. 
The form of pressure is the same as the commonly used virial but multiplied by a localisation function $\StressIKoneinline = \StressVIRIALinline \vartheta_i$. Strictly speaking, the virial form is only valid for an entire periodic MD simulation (i.e. no localisation with a Dirac delta of $\vartheta$ function) but is widely used locally in bins due to is simplicity and implementation in codes like LAMMPS \citep{LAMMPS}. 

The virial pressure arbitrarily splits the contribution into half per atom, while the Volume Average (VA) pressure \eq{VA_stress}) distributes these along the interaction path. As a result, the VA requires us to define an interaction path between two atoms, something which is non-trivial given the many-body nature of the MACE potential. 
Instead, we can express \eq{Config_pressure} in terms of stress across the surface of the volume, here assumed to be a cuboid for simplicity, by simply evaluating the derivative of $\vartheta_{\ell}$.
This gives a molecular version of the divergence theorem \citep{Smith_et_al12}, 
 \begin{align}
\int_V \boldsymbol{\nabla} \cdot \Stresstype{}^C  dV  & =  - \frac{1}{2}\displaystyle\sum_{i,j}^{N} \bigg\langle   	\boldsymbol{f}_{ij} \cdot  \oint \frac{\partial  \vartheta_{\ell} }{\partial \boldsymbol{r}}   d\boldsymbol{\ell}  ; \textit{f}  \bigg\rangle 
\nonumber \\
& =- \frac{1}{4} \displaystyle\sum_{\beta \in \{\text{faces}\}} \displaystyle\sum_{i,j}^{N} \bigg\langle  \boldsymbol{f}_{ij} d_{\beta_{ij}} S_{\beta ij}   ; \textit{f}  \bigg\rangle
\nonumber \\
& = \displaystyle\sum_{\beta \in \{\text{faces}\}} \int_{S_{\beta}} \Stresstype{}^C \cdot d\textbf{S}_{\beta}.
\label{CV_volume}
 \end{align}
On a CV this is then localised to each face, here a cuboid, with six faces $\beta = \{x^+$, $x^-$, $y^+$, $y^-$, $z^+$, $x^-$\}. 
The result is an expression which is non-zero \textit{only} when the interaction between two atoms passes over a face with $\; d_{\beta_{ij}} = sgn(\beta - \beta^d_{j})- sgn(\beta - \beta^d_{i})$, ensuring the two atoms are on either side of an infinite plane extending from the CV face.
Here $\beta^d$ denotes the coordinate direction of $\beta$ (e.g. for $\beta = x^+$ we have $\beta^d = x$).
The function $S_{\beta ij}$ then identifies if this crossing is localised to a face, with form $S_{\beta ij}=\prod^{\alpha \in \{\text{x,y,z}\}}_{\;\;\;\alpha \ne \beta^d} [ H(\alpha^+ - \alpha_i) -  H(\alpha^- - \alpha_i)]$ a product over the two remaining dimensions excluding $\beta^d$. 

Comparing \eq{varthetaij} and \eq{CV_volume} shows the difference of two $\vartheta$ functions is directly equal to the sum of all six surface crossings,
 \begin{align}
\vartheta_i - \vartheta_j = \oint \frac{\partial  \vartheta_{\ell} }{\partial \boldsymbol{r}} d\boldsymbol{\ell}  =  \frac{1}{2}\displaystyle\sum_{\beta \in \{\text{faces}\}} d_{\beta_{ij}} S_{\beta ij}
\label{vartheta_to_surface}
 \end{align}
which is consistent with the previous interpretation of only surface crossings contributing to momentum change in the volume.
Taking a given face, say $\beta = z^+$ the dot product of the rank 2 stress tensor $\Stresstype{}^C$, with unit vector for face $z^+$,
\begin{align}
\int_{S_{z^+}} \Stresstype{}^C \cdot d\textbf{S}_{z^+} = \StressCV{}_{\!\!{z^+}}^{\!\!C} \Delta A_{z^+}^{CV},
\label{Surface_dot}
\end{align}
where we have implicitly assumed the value of pressure on the face is a constant.
The bold notation on the right of \eq{Surface_dot} for $\StressCV{}_{\!\!{z^+}}^{\!\!C}$ denotes surface pressure, a vector with 3 components on one surface (a rank 1 tensor). 
This becomes the rank 2 stress tensor when considering a tetrahedron with 3 surfaces as used by Cauchy in his original stress tensor definition.
For a cuboid in space, each face has three components of stress, which results in $18$ independent components over the total control surface. 
The control volume pressure on the $z^+$ is therefore,
\begin{align}
	\StressCV{}_{\!\!{z^+}}^{\!\!C}  
		 = \frac{1}{4 \Delta A_{z^+}^{CV}} \displaystyle\sum_{i,j}^{N} \bigg\langle  \boldsymbol{f}_{ij} d_{z^+_{ij}}S_{z^+_{ij}}   ; \textit{f}  \bigg\rangle,
\label{Surface_pressure}
 \end{align}
that is, the pressure on surface of area $\Delta A_{z^+}^{CV}$. 
The signum functions $d_{z^+_{ij}}= sgn(z^+ - z_j)- sgn(z^+ - z_i)$ are only non zero if molecule $i$ and $j$ are on different sides of a plane at $z^+$  while the $S_{zij}^+=\prod_{\alpha \in \{\text{x,y}\}} [ H(\alpha^+ - \alpha_i) -  H(\alpha^- - \alpha_i)]$ term specifies the point of intersection of the line is located on a region of the $z^+$ plane, in \eq{Surface_pressure} the surface of the cuboid. 
The surface is the localised form of the pressure tensor considered by \citet{Han_Lee} but applied to the six cubic faces bounding a volume. 
Surface localisation to a cube is not trivial to define for the MACE potential, so we instead consider a series of slabs matching the width ($L_x$) and breadth ($L_y$) of the domain.
This is achieved by setting $S_{zij}^+=1$, so the surface located at $z^+$ spans the entire domain as an infinite plane in \eq{CV_pressure}, recovering the Method of Planes formulation of the pressure \citep{Todd_et_al_95}, 
\begin{align}
	\StressMOP{}_{\!\!\!\!\!z^+} = \frac{1}{\Delta A_{z^+}} \displaystyle
		\sum_{i = 1 }^{N} \bigg\langle \frac{{\boldsymbol{p}}_i {p}_{iz}}{m_i} \delta(z_i-z^+) ; \textit{f}  \bigg\rangle 	\nonumber \\ 
+ \frac{1}{4\Delta A_{z^+}} \displaystyle\sum_{i,j}^{N} \bigg\langle \boldsymbol{f}_{ij} d_{z^+_{ij}} \!  ; \textit{f}  \bigg\rangle. \;\;\;\;\;
\label{CV_pressure}
\end{align}
This is the well known form of the method of planes pressure \citep{Todd_et_al_95}, which is a vector for the 3 components of stress on the $z$ plane, $P_{zx}$, $P_{zy}$  and $P_{zz}$.
Any two planes bounding a region in periodic space can be considered to form a control volume, so the molecules between satisfy the conservation \citep{Smith_et_al12} with time evolution of momentum from \eq{IK_timeevo} given by,
\begin{align}
	\frac{\partial }{\partial t} \int_V \rho  \boldsymbol{u} dV  =\frac{\partial }{\partial t}  \displaystyle\sum_{i=1}^{N}  \bigg\langle \boldsymbol{p}_i \vartheta_i ; \textit{f} \bigg\rangle  = \StressMOP{}_{\!\!\!\!\!z^+} - \StressMOP{}_{\!\!\!\!\!z^-}.
\label{MOP_pressure}
\end{align}
This equation is also valid instantaneously, i.e. without the ensemble average denoted by $\big\langle ; \textit{f} \big\rangle$, a result shown in past work for pairwise potentials \citep{Smith_et_al12} and extended here to ML potentials in Section \ref{sec:results}.
Note the minor sleight of hand in \eq{MOP_pressure} swapping to laboratory momentum $\overline{\boldsymbol{p}}_i \to \boldsymbol{p}_i$ in \eq{CV_pressure}, as the streaming velocity is zero in this system.
This simplifies \eq{MOP_pressure} and removes the need for convective $\rho \boldsymbol{u}\boldsymbol{u}$ terms.
These can be included if considering cases with a flow, taking care to include the "plane peculiar" streaming velocity evaluated at the plane, not the particle itself \citep{Todd_et_al_95, Smith_et_al19}. 
It is the conservative property of \eq{MOP_pressure} which will be tested using the MACE potentials in this work.
Given the path is non unique, and even the interaction force between atoms is arbitrary, \eq{CV_pressure} has the advantage of only requiring that the atoms be either side of a plane to get the pressure.
This is unambiguous for both definitions of $\boldsymbol{f}_{ij}$ and $\widetilde{\boldsymbol{f}}_{ij}$ in this work as we require interactions be counted only if $i$ is one side of a plane and $j$ the other \footnote{Note however some work introduces a molecular centre of mass in the surface crossing calculation of angular and torsion bonds \citep{Heinz2007} which would complicate this. We treat all atoms as pairwise using MACE so only $i$ and $j$ position are relevant.}.
It is shown in section \ref{sec:results} that both forms of $\boldsymbol{f}_{ij}$ and $\widetilde{\boldsymbol{f}}_{ij}$ give the same momentum balance.
This pressure can then be checked to ensure it satisfies conservation of momentum in \eq{MOP_pressure}, providing validation that the many-body MACE pressure we are measuring is exactly consistent with the resulting dynamics of the molecules.
We show in section \ref{sec:results} that the virial stress of \eq{virial} does not satisfy the momentum balance condition near a wall, a well-know result in the literature \citep{Todd_et_al_95, heyes2011equivalence} even in the steady state. 
However, we cannot evaluate the volume average unambiguously as we do not have the intermolecular interaction path, and even if we could, it cannot be linked to the time evolution in an exact way \citep{Smith_et_al12}.

\subsection{Energy Equation}
\label{sec:energy}
The momentum balance on a control volume, obtained in the previous section, uses a force definition considering the derivative of the entire system energy landscape $U(\{r_{ij}\})$ as a function of pairwise network graph connections.
However, to get the energy change in a given volume, the concept of potential energy per particle has to be introduced so we can quantify how much energy is in a volume.
The energy at a point from \citet{Irving_Kirkwood} is given by,
\begin{align}
\rho\mathcal{E} \define \displaystyle\sum_{i=1}^{N} \bigg\langle e_i  \delta(\boldsymbol{r}_i-\boldsymbol{r}) ; \textit{f} \bigg\rangle,
\label{Energydensity}
\end{align}
where the energy per atom is $e_i = \boldsymbol{p}_i^2/2m_i + U_i$ and the sum of all atoms gives total energy $U = \sum_{i=1}^N U_i$.
Now to get the evolution of energy \citet{Irving_Kirkwood} use \eq{IK_timeevo} with $\alpha = \sum e_i \delta(\boldsymbol{r} - \boldsymbol{r}_i)$,
\begin{align}
\frac{\partial }{\partial t} \rho\mathcal{E} & = \frac{\partial }{\partial t}  \displaystyle\sum_{i=1}^{N}  \bigg\langle e_i  \delta(\boldsymbol{r}_i-\boldsymbol{r}) ; \textit{f} \bigg\rangle  \nonumber \\
& = \displaystyle\sum_{i = 1 }^{N} \bigg\langle  \underbrace{\left[ \boldsymbol{F}_i \cdot \frac{\partial e_i}{\partial \boldsymbol{p}_i} +\frac{\partial e_i}{\partial \boldsymbol{r}_i} \cdot  \frac{\boldsymbol{p}_i}{m_i} \right]}_{\mathcal{A}_i}  \delta(\boldsymbol{r} - \boldsymbol{r}_i)) 
\nonumber \\
& \;\;\;\;\;\;\;\;\;\;\;\;\; - \frac{\partial}{\partial \boldsymbol{r}} \cdot e_i \frac{\boldsymbol{p_i}}{m_i}  \delta(\boldsymbol{r} - \boldsymbol{r}_i))  ;\textit{f} \bigg\rangle, 
\label{energy_timeevo}
 \end{align}
The quantity on the final line is the advection of energy and does not include interactions. We focus on the quantity $\mathcal{A}_i$ in the bracket, with $\partial e_i / \partial \boldsymbol{p}_i = \boldsymbol{p}_i / m_i$ and $\partial e_i / \partial \boldsymbol{r}_i = \partial U_i / \partial \boldsymbol{r}_i$, this becomes,
\begin{align}
\mathcal{A}_i & = \boldsymbol{F}_i \cdot \frac{\boldsymbol{p_i}}{m_i} +  \frac{\partial U_i}{\partial \boldsymbol{r}_i} \cdot  \frac{\boldsymbol{p}_i}{m_i} 
\nonumber \\
& = -\frac{\partial U}{\partial \boldsymbol{r}_i} \cdot \frac{\boldsymbol{p_i}}{m_i} + \sum_{j =1}^N \frac{\partial U_i}{\partial \boldsymbol{r}_j} \cdot   \frac{\partial \boldsymbol{r}_j}{\partial \boldsymbol{r}_i} \cdot  \frac{\partial \boldsymbol{r}_i}{\partial t} \nonumber \\
 & = \sum_{j =1}^N \frac{\partial U_i}{\partial \boldsymbol{r}_j} \cdot  \frac{\boldsymbol{p_j}}{m_i} -\frac{\partial \sum_{j =1}^N U_j }{\partial \boldsymbol{r}_i}  \cdot \frac{\boldsymbol{p_i}}{m_i}. 
\label{EQ_A}
\end{align}
Where the following definitions are also applied $\boldsymbol{p}_i/m_i = \partial \boldsymbol{r}_i /\partial t$, $\boldsymbol{F}_i = - \partial U / \partial \boldsymbol{r}_i$ and $U = \sum_{i} U_i $.
Taking the sum over all $i$ and multiplying \eq{EQ_A} by the Dirac delta function gives $\sum_i \mathcal{A}_i \delta (\boldsymbol{r} - \boldsymbol{r}_i)$ which becomes,
\begin{align}
\displaystyle\sum_{i = 1 }^{N} \sum_{j =1}^N \bigg\langle  \frac{\partial U_i}{\partial \boldsymbol{r}_j} \cdot  \frac{\boldsymbol{p_j}}{m_i} \delta(\boldsymbol{r} - \boldsymbol{r}_i))   -\frac{\partial U_j}{\partial \boldsymbol{r}_i}   \cdot \frac{\boldsymbol{p_i}}{m_i} \delta(\boldsymbol{r} - \boldsymbol{r}_i))  ;\textit{f} \bigg\rangle, 
\nonumber \\
 =\displaystyle\sum_{i,j}^{N} \bigg\langle  \frac{\partial U_i}{\partial \boldsymbol{r}_j} \cdot  \frac{\boldsymbol{p_j}}{m_i} \delta(\boldsymbol{r} - \boldsymbol{r}_i))   -\frac{\partial U_i}{\partial \boldsymbol{r}_j}   \cdot \frac{\boldsymbol{p_j}}{m_j} \delta(\boldsymbol{r} - \boldsymbol{r}_j))  ;\textit{f} \bigg\rangle, \nonumber \\
 =\displaystyle\sum_{i,j}^{N}  \bigg\langle  \frac{\partial U_i}{\partial \boldsymbol{r}_j} \cdot  \frac{\boldsymbol{p_j}}{m_i} \left[ \delta(\boldsymbol{r} - \boldsymbol{r}_i))   -\delta(\boldsymbol{r} - \boldsymbol{r}_j)) \right]  ;\textit{f} \bigg\rangle,
 \end{align}
This can be written as the MoP stress using the same process as the momentum equation, integrating over a volume and evaluating the derivative (or simply use \eq{vartheta_to_surface}).
For a given direction and surface, say $z^+$ the MoP energy equation is then,
\begin{align}
	\HeatfluxMOP{\!\!\!}_{z^+}^{K} = &\frac{1}{\Delta A_{z^+}} \displaystyle
		\sum_{i = 1 }^{N} \bigg\langle \frac{e_i {p}_{iz}}{m_i} \delta(z_i-z^+) ; \textit{f}  \bigg\rangle 	\nonumber \\
\HeatfluxMOP{\!\!\!}_{z^+}^{C} = &\frac{1}{4\Delta A_{z^+}} \displaystyle\sum_{i,j}^{N} \bigg\langle \frac{\partial U_i}{\partial \boldsymbol{r}_j} \cdot  \frac{\boldsymbol{p_j}}{m_i}  d_{z^{+}_{ij}} \!  ; \textit{f}  \bigg\rangle. 
\label{MOP_energy}
\end{align}
So given $\HeatfluxMOP{\!\!\!}_{z^+} = \HeatfluxMOP{\!\!\!}_{z^+}^{K} + \HeatfluxMOP{\!\!\!}_{z^+}^{C}$, the energy conservation of a control volume between two $z$ planes can be written as,
\begin{align}
	\frac{\partial }{\partial t} \int_V \rho\mathcal{E} dV & = \frac{\partial }{\partial t}  \displaystyle\sum_{i=1}^{N}  \bigg\langle e_i  \vartheta_i ; \textit{f} \bigg\rangle   = \HeatfluxMOP{\!\!\!}_{z^+}- \HeatfluxMOP{\!\!\!}_{z^-}
\label{energy_conservation}
\end{align}
which will not be satisfied to machine precision, as with momentum, but to comparable accuracy to the systems energy conservation.
It is useful to review the pseudocode that can obtain the configurational part of the heat flux,
\begin{verbatim}

#Loop needed over all energies Ui per particle
for i in Natoms:
  dUidrj[i,:,:] = autograd(node_energy[i], r)

#Sum over all i and j for a 
#given plane located at zplane
for i in Natoms:
  for j in Natoms:
    #Obtain work done to be used in MoP term
    fijvi = -( dUidrj[i,j,0]*v[i,0]
              +dUidrj[i,j,1]*v[i,1]
              +dUidrj[i,j,2]*v[i,2])
    dzij = ( sgn(zplane - r[j,2])
            -sgn(zplane - r[i,2]))
    MoPpower_c += 0.5*fijvi*dzij
\end{verbatim}
It is possible to observe from this implementation that the operation will be computationally expensive. For each atom i the full back-propagation operation over the entire network must be used to get each $\partial U_i / \partial \boldsymbol{r}_j$.
A more efficient version is given in \citet{Langer2023heatflux} based on defining a position variable outside of the computational graph.
This approach appears to be valid only for a whole domain so is unlikely to be possible to extend to the MoP pressure in the form presented here.
One potential way forward using the control volume conservation is sketched out in the appendix \ref{sec:heatflux}

The difference in the form of pairwise force $\partial U_i / \partial \boldsymbol{r}_j$ obtained in \eq{MOP_energy} compared to the pair based $\partial U / \partial \boldsymbol{r}_{ij}$ in \eq{Enforce_Newton3rd} can be understood in terms of how they distribute the "work".
The use of $\partial U_i / \partial \boldsymbol{r}_j$ assigns work to the specific atom while $\partial U / \partial \boldsymbol{r}_{ij}$ assigns this work to the graph edge, as a result it fails local control volume energy conservation, because it spatially misattributes where the energy change is happening. 
It is worth expressing this naive approach to getting the heat flux in terms of graph edges $\partial U / \partial \boldsymbol{r}_{ij}$; with the pairwise forces of \eq{Enforce_Newton3rd} inserted into the commonly used pairwise MoP form from the literature \citep{Han_Lee, Smith_et_al19},
\begin{align}
	\widetilde{\HeatfluxMOP{\!\!\!}^C}_{\!\!\!z^+} & = \frac{1}{4\Delta A_{z^+}} \displaystyle\sum_{i,j}^{N} \bigg\langle \boldsymbol{f}_{ij} \cdot  \frac{\boldsymbol{p_j}}{m_i}  d_{z^{\!+}_{ij}}\!  ; \textit{f}  \bigg\rangle \nonumber \\
& = \frac{1}{4\Delta A_{z^+}} \displaystyle\sum_{i,j}^{N} \bigg\langle \left( \frac{\partial U}{\partial \boldsymbol{r}_{ij}} - \frac{\partial U}{\partial \boldsymbol{r}_{ji}}\right) \cdot  \frac{\boldsymbol{p_j}}{m_i}  d_{z^{\!+}_{ij}} \!  ; \textit{f}  \bigg\rangle 
\label{MOP_energy_pairwise}
\end{align}
This equation gives an error as shown in section \ref{sec:results}, but is much more computationally efficient given only a single back-propagation step is required to obtain $\partial U / \partial \boldsymbol{r}_{ij}$.
How important this error is for a given local heat flux measurement is something which should be carefully checked for each problem being studied.

\begin{figure}
\includegraphics[width=0.45\textwidth]{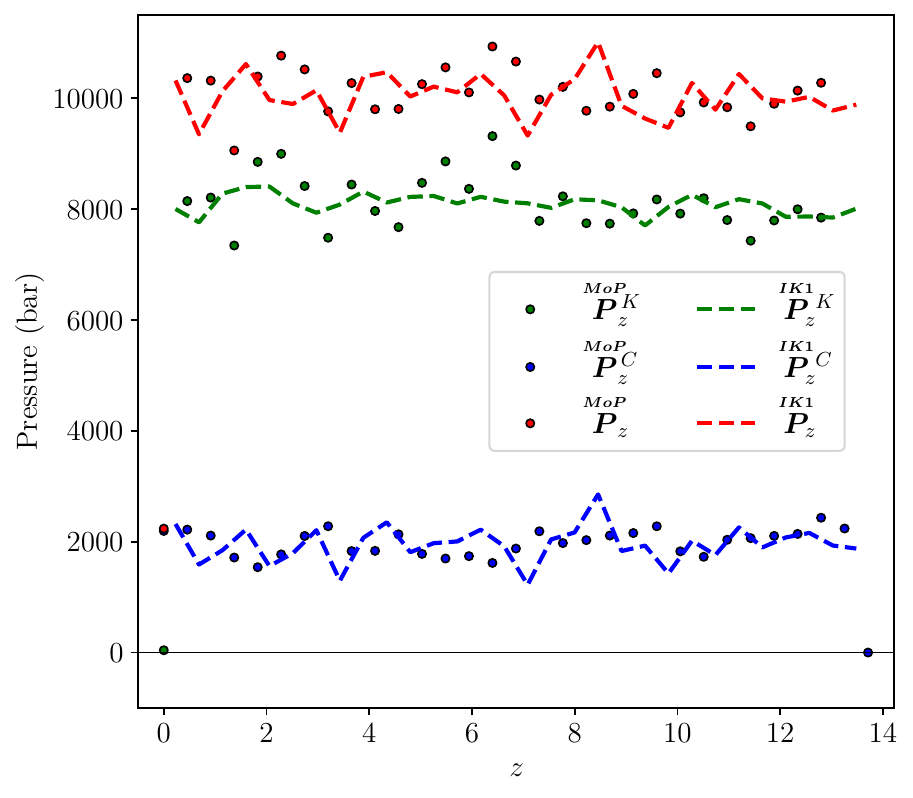}
\caption{Pressure in a pure water NPT simulation controlled to 10,000 bars comparing virial pressure ($\StressIKoneinline$ from \eq{virial}) to Method of Planes over 30 bins and 31 planes respectively. Plots are an average over 30,000 samples in time taken every 10 timesteps steps from MD simulations.}  
\label{fig:MOP_NPT}
\end{figure}

\begin{figure*}
\includegraphics[width=0.95\textwidth]{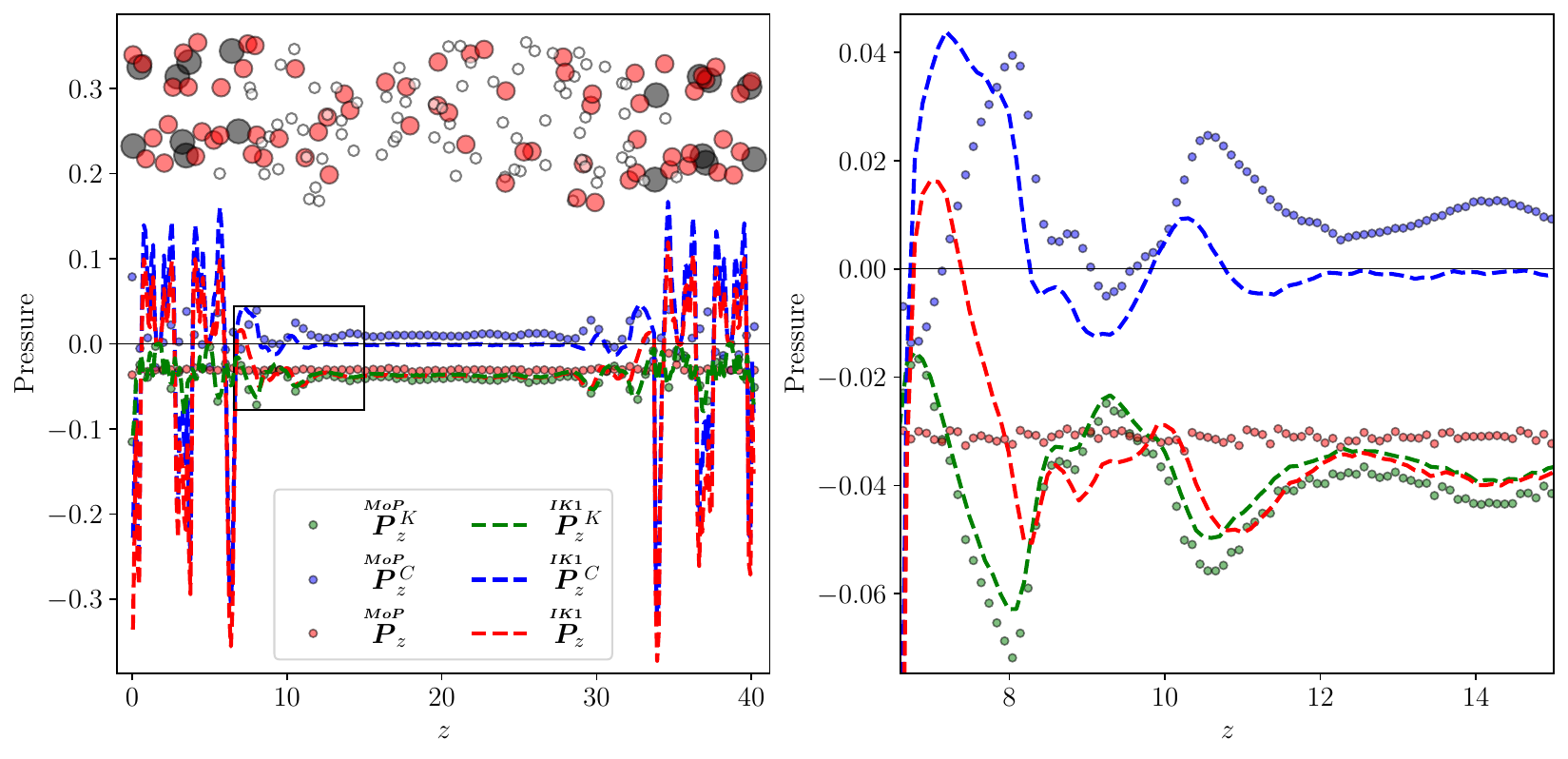}
\put(-446,33){$a)$}
\put(-203,33){$b)$}
\caption{Plot of spatially localised virial pressure ($\StressIKoneinline$ from \eq{virial}) compared to Method of Planes over 400 bins and 401 planes respectively. The full channel is shown in $a)$ with MoP plotted every 5th bin below the molecular locations from a trajectory snapshot (Zr black, oxygen red and hydrogen white) demonstrating how pressure changes from the solid ZrO$_2$ to the liquid H$_2$O regions. The zoomed in near-wall region is shown in $b)$, where the colours/legend is consistent in both plots. Plots are an average over 30,000 samples in time taken every 10 timesteps steps from MD simulations, with symmetry assumed to improve statistics in $b)$, to 60,000 samples. The kinetic pressure and configurational pressure must sum to a constant for $\partial P_{zz} / \partial z = 0$ to be valid, seen as a constant value for the red point with MoP but not for the IK1/virial sum (dashed red line) as the measured pressure is biased by being assigned to the location of the molecules.}  
\label{fig:MOP_balance}
\end{figure*}

\section{Methodology}
\label{sec:methodology}

The focus of this work is to demonstrate that the surface form of pressure is valid in a non equilibrium molecular dynamics simulation with the MACE potential.
Given the potential range of models that could be studied with MACE, we choose water at an interface with Zirconium Oxide (zirconia); a well-studied oxide with application in thermal barrier coatings, catalysis and to prevent corrosion in zirconium alloys \citep{Yang_et_al2021}.
The choice of studied material is somewhat arbitrary here as the focus is on obtaining stresses, and MACE requires no differing assumption when modelling any of the 89 elements in its training data.

An initial NPT test is used to ensure the pressures given by the MoP and Virial are consistent and correct, a pure water system is run with $80$ molecules in a periodic domain starting with a size of $13.61 \times  13.24 \times 13.24$.
The simulation uses the Atomic Simulation Environment (ASE) \citep{ase-paper} with the MACE potential calculator.
An Inhomogeneous Berendsen Barostat is used, which varies the domain extents only in the $L_x$ and $L_y$ directions so the 30 bins and 31 planes in the z direction always cover the domain.
As a result the division by area in \eq{CV_pressure} and volume in \eq{virial} varies each time based on the Berendsen rescaling of $L_x$ and $L_y$.
As water is mostly incompressible, large fluctuations are observed in pressure so the NPT ensemble is run with a high pressure setpoint of $P_{set} = 10,000$bar to ensure good signal to noise ratio.
The time constant for Berendsen pressure is $1000fs$ with compressibility set to $4.57 \times 10^5 bar^{-1}$, corresponding to a bulk modulus of approximately $2.2 \times 10^4 bar$
A high temperature of $600K$ is also used to ensure the molecular motion remains liquid, with a thermostat time constant of $100fs$.
The simulation is run over $20,000$ timesteps of $\Delta t=0.5$ sampled at $10$ timestep intervals, with the first $2,000$ timesteps used to equilibrate the system and then the remaining used for statistical averaging.

For the remaining simulations in this work, a water ZrO$_2$ system is studied. 
The simulation setup is taken from \citet{Yang_et_al2021} which studied an ab initio molecular dynamics simulations of an interface between monoclinic phase ZrO$_2$ with the (1$\bar{1}$1) surface orientation at the interface to the liquid water.
This is shown in Figure \ref{fig:domain} highlighting the dimensions of the solid and liquid regions.
This particular surface was chosen as it is reported to have the lowest energy orientation \citep{Christensen_et_al98}. The simulation cell contained 96 formula units of ZrO$_2$, as in \citet{Yang_et_al2021} but with the number of water molecules doubled to give a larger fluid region. The simulation cell is triclinic, to accommodate the ZrO$_2$ cell, with dimensions $13.6\AA \times 13.6\AA \times 40.3\AA$ in $x$, $y$ and $z$ respectively. The solid walls are about $7\AA$ at the top and bottom, with periodic boundaries connecting all sides.
There is no need for fixing or tethering of the walls, as the use of ZrO$_2$ gives a stable interface with the water.
No fluid flow or thermal gradient is applied in these simulations, which removes the added complexity of streaming velocity in the pressure and heat flux measurements.
Instead, the NEMD behaviour of the system is explored by taking very short timescales in local regions, where the assumption of global equilibrium is not valid and the balance of time evolving and flux terms in the control volume equations validate the MoP pressure away from equilibrium.
The system is also inhomogeneous with a liquid-solid interface, providing spatial variation which is sufficient to explore the known limitation of the virial form of pressure.
The domain was split into 400 control volume slabs, each of size $\Delta z=0.1\AA$, with the momentum and energy in the volume collected along with the stress and heat flux on the top and bottom planes of each volume.
Unless otherwise stated, all quantities in this work are given in the standard units for ASE, where $amu=eV=$\AA$=K=1$ and times in the ASE time unit are $t=\AA\sqrt{amu/eV}$ which means one ASE time unit is approximately $10.2fs$.

Atomic interactions were described using the MACE machine-learned interatomic potential \citep{batatia2022mace}. A range of models were tested, including \texttt{mace-mpa-0-medium} and \texttt{MACE-MATPES-PBE-0} run with a customised branch of the Github MACE code edited to return pairwise forces \footnote{branched at commit e2842aadd15a57580e8564b050bd6b4eee852b7d with full code and all changes at \href{https://github.com/edwardsmith999/mace/}{https://github.com/edwardsmith999/mace/} for reproducibility. The ASE code to do this was added to the main MACE repository by the author and this functionality has been included in the main MACE releases as edge\_forces from version 0.3.13 onwards. An example of using MACE versions later than 0.3.13 is also included in the code provided with this publication.}. The model was run on a CUDA-enabled GPU (NVIDIA 4060Ti) using the NVIDIA cuEquivariance library for acceleration \citep{cuEquivariance}, run with pytorch 2.8, CUDA kernel 12.6 and CUDA Version 560.35.03. The calculation of surface crossings for the MoP calculation is accelerated using Numba \citep{Numba}. 

The MACE-MATPES-PBE-0 version of MACE has no Hubbard +U correction which matches the setup of \citet{Yang_et_al2021}. Behaviour such as disassociation of water is observed near the surface. Such behaviour is highly non-trivial to capture with classical models, but for DFT and MACE systems, the water molecules are held together purely by the MACE forces requiring no explicit bonds. The electronic structure calculations underpinning MACE training use the generalised gradient approximation (GGA) with the Perdew-Burke-Ernzerhof (PBE) functional, a ground state and relatively simple form of DFT. However, given the rapid evolution of both MACE and general ML style potentials, it is expected future potentials will continue to change as the database and training become more extensive. This is especially true as MACE-MP-0 is designed to be a foundation model which is meant to be fine tuned using DFT to explicitly cover specific case. These include improving the cross interactions between the ZrO$_2$ wall and water atoms for example.
Although \citet{batatia2023foundation} show water interface demonstrate reasonable results, recent work shows MACE does lose accuracy for interfaces \citep{Focassio_et_al2025}. However, \citet{Focassio_et_al2025} also show that fine tuning can improve a foundation models much faster than starting training from scratch.
The model does not include dispersion terms and long-range electrostatics.
However, the classical MoP formulation has not been adapted for long-range forces to the author's knowledge, so this limitation would remain for QM systems.
The use of the MoP together with these potentials will therefore not accurately reproduce a real system where electrostatics are important, and an extension of the MoP needs to be developed for these systems.

\citet{Yang_et_al2021} ran water at about 30K above room temperature to compensate for the overstructuring common in DFT simulations. 
However, in the MACE simulation at 330K temperature with an NVE ensemble, considerable over structuring was still observed in the water.
This shows minimal flow, instead forming a kind of glassy crystal, even with the addition of D3 dispersion corrections \citep{D3_correction}. 
To force more fluid like behaviour, the simulations were performed at a temperature of 500K, which was seen to provide extensive movement of the water atoms. 
This elevated temperature was specifically selected to help correct for the known over-binding error of density functional theory (DFT) when simulating bulk water \citep{Lee_et_al94} and ensure we can test the MoP equations for water in the liquid phase. 
This also corresponds to the lower end of typical temperatures used in nuclear reactors \citep{Yang_et_al2021}, providing a physical justification for the studied system.
A Nose-Hoover thermostat was used to elevate the entire system as an NVT for an equilibration period.
The simulation was then restarted as an NVE with all results collected during a constant energy run for the production phase. 
For the runs to test the spatial variation of pressure, sufficient sampling is required so longer runs of ($300,000$ timesteps at $\Delta t=0.5$) with samples every 10 timesteps collected giving $30,000$ samples for the pressure.
For the control volume conservation tests, only very short runs of $40$ time units were required with 80 timesteps at $\Delta t=0.5$ for momentum and 320 timesteps at $\Delta t=0.125$ for energy.

\section{Results}
\label{sec:results}

We start with the NPT simulation barostatted to a setpoint of $10,000$bar, which tests that the MoP pressure has the expected value and is consistent with the local IK1 or its average over the whole domain giving the global virial pressure.
Figure \ref{fig:MOP_NPT} shows the MoP pressure compared to the IK1 pressure. 
This shows good agreement between the MoP measurements and IK1 (virial) with both expressed in real units. 
The measurements on planes at the domain top and bottom show values of zero, attributed to an issue with crossings not being correctly counted over periodic boundaries.

Next, the water ZrO$_2$ system is simulated.
It has been documented in the literature \citep{Todd_et_al_95, heyes2011equivalence} that using the virial form of pressure locally (IK1) fails to give a flat profile near walls.
This must be an error, as the condition for mechanical equilibrium $\nabla \cdot \boldsymbol{P} = 0$ is violated.
This in turn implies the locally applied virial (IK1) form of pressure cannot be correct as the result of a non-zero gradient violates mechanical equilibrium and results in a net flow, something not possible for static fluid between two surfaces.
These anomalous peaks in pressure observed in the virial are simply a consequence of assigning the pressure at the location of the molecules, which tend to stack showing density peaks near walls.
These artefacts appear due to the virial assumption of a half-half split of pressure contribution from an interactions, assigning these disproportional to the location of molecules. 
Instead, mechanical pressure should be defined at an arbitrary plane in space, as in the original definition of Cauchy.
The MoP kinetic (green points) and configurational (red points) pressure in Figure \ref{fig:MOP_balance} can be seen to have a symmetry, with the same shape as a function of $z$.
These represent the variations in pressure due to the liquid structure near the wall.
As kinetic values become more negative, the configurational part becomes positive by the same amount.
These balance exactly and the total pressure shown as red points in Figure \ref{fig:MOP_balance} is a constant value.
This constant value of total pressure is the only physically meaningful value in a simulation with no flow. 
This is an important validation as it ensures the form of pressure satisfies $\nabla \cdot \boldsymbol{P} = 0$, despite the arbitrary definition of a pairwise intermolecular force $\boldsymbol{f}_{ij}$ from the multi-body MACE potential.

\begin{figure*}
\includegraphics[width=0.95\textwidth]{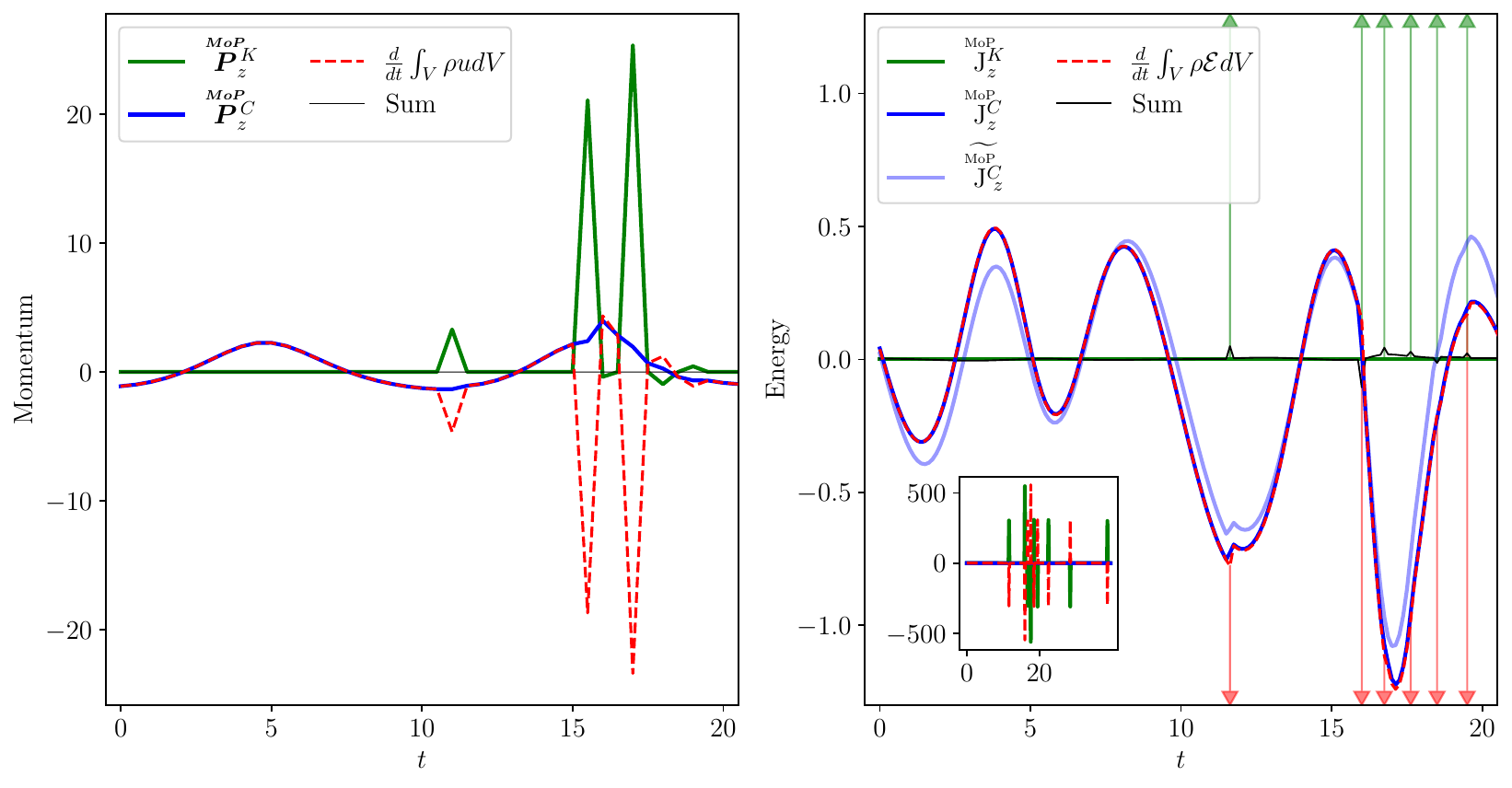}
\put(-445,33){$a)$}
\put(-203,33){$b)$}
\caption{Control volume conservation plots, $a)$ is momentum conservation for a simulation with $\Delta t = 0.5$, where the measured configurational forces and momentum fluxes are equal to momentum change, shown by the sum which is zero to machine precision. $b)$ Energy, at smaller timestep $\Delta t = 0.125$ so crossings are shown as arrows and both pairwise force are compared: with $\widetilde{\Heatfluxtype{}^C}_{\!\!\!z}$  based on $\left[\frac{dU}{d\boldsymbol{r}_{ij}}  -\frac{dU}{d\boldsymbol{r}_{ji}}  \right] \cdot \frac{\boldsymbol{p}_j}{m_j}$ and $\Heatfluxtype{}^C_{z}$ based on $\frac{dU_i}{d\boldsymbol{r}_j} \cdot \frac{\boldsymbol{p}_j}{m_j}$. Sum is based on the more accurate $\frac{dU_i}{d\boldsymbol{r}_j} \cdot \frac{\boldsymbol{p}_j}{m_j}$ form. Insert shows full scale, highlighting molecular surface crossings with magnitudes$\to \infty$ as $\Delta t \to 0$. The plot is for volume number 211, roughly in the middle of the channel, although similar plots can be obtained for any volume. 
}  
\label{fig:CV_trace}
\end{figure*}

The other check that the measured pressure is meaningful is demonstrated in Figure \ref{fig:CV_trace}$a)$. This control volume check ensures that $d/dt \int_v \rho \boldsymbol{u} dV = \oint \boldsymbol{P} \cdot d\textbf{S} $, that is the forces $\boldsymbol{f}_{ij}$ which are summed over a plane to give $\boldsymbol{P}$ in \eq{CV_pressure} exactly determine the change of momentum of the molecules. This is shown to machine precision in Figure \ref{fig:CV_trace}, where the sum of the kinetic spikes as molecules cross $\StressMOP{}^K$ plus the total summed $\boldsymbol{f}_{ij}$ contributions in $\StressMOP{}^C$ from all molecules are equal to the time evolution.
Again, the arbitrary nature of the pairwise force mean this check is useful, showing the measured MoP pressure satisfy the continuum momentum equations exactly.
Variants of this force decomposition, including $\boldsymbol{f}_{ij} = -\partial U_i/ \partial \boldsymbol{r}_{j}$ and even just $\boldsymbol{f}_{ij} = -\partial U/ \partial \boldsymbol{r}_{ij}$ without corresponding transpose to enforce Newton's 3rd law were tested and demonstrate identical values of $\StressMOP{}^C$.
This suggests interactions on a plane removes some of the ambiguity introduced by the non-unique pairwise forces, and that these varying forms all ensure momentum conservation.

The two definitions of the energy, namely the pairwise-force formulation from \eq{MOP_energy_pairwise} and the expression derived by \citet{Langer2023heatflux} that leads to the MACE energy conservation relation in \eq{MOP_energy}, were evaluated in Figure \ref{fig:CV_trace}$b)$. 
For notational conciseness, we drop the MoP superscript in the text and captions, $\HeatfluxMOP{\!\!\!} \to \Heatfluxtype$, as only MoP heat flux is considered.
The results show clearly that the formulation in \eq{MOP_energy}, which employs $\partial U_i / \partial \boldsymbol{r}_{i}$, yields substantially improved energy conservation. 
These energy traces were generated using a smaller timestep than in the momentum example, because energy conservation in molecular dynamics is intrinsically approximate, the integrator conserves a shadow Hamiltonian \citep{Hammonds_Heyes2021}, with an error that decreases as the timestep is reduced. 
For the expression in \eq{MOP_energy}, this error decreases with timestep, whereas for the pairwise formulation in \eq{MOP_energy_pairwise} the error persists even in the small-timestep limit, indicating a systematic issue in using $\boldsymbol{f}_{ij} = \partial U/ \partial \boldsymbol{r}_{ij}-\partial U/ \partial \boldsymbol{r}_{ji}$ in the energy conservation equation.
The error between $\widetilde{\Heatfluxtype{}^C}_{\!\!\!z^+}$ and ${\Heatfluxtype{}^C}_{\!\!\!z^+}$ does not follow a clear pattern, with sometimes under prediction ($t \sim 4$), sometimes over prediction ($t \sim 19$) and even close agreement (around times 6-7 or 14).
This apparently random error is likely due to the somewhat arbitrary nature of the energy assignment used, with $\widetilde{\Heatfluxtype{}^C}_{\!\!\!z^+}$ due to pairwise interactions or graph edges giving a varying difference from the particle based $\partial U_i / \partial \boldsymbol{r}_{i}$ in $\widetilde{\Heatfluxtype{}^C}_{\!\!\!z^+}$.
Sometimes by chance these will be consistent and at other times have a larger or smaller value.

The energy change in a control volume is due to the work done on the atoms from outside the cell, with slow changes due to forces and sudden peaks due to energy carried into or out of the volume by atomic surface crossings.
These peaks have magnitudes around 500, as can be seen from the inset in Figure \ref{fig:CV_trace}$b)$, so the errors in energy conservation observed during crossings are relatively small.
These may also be a consequence of the calculation of the work done being assigned to bins based on atom position at the start of the timestep, whereas in practice, the crossing occurs at some point in the timestep, so the work done will be split between volumes. 
The agreement between work done and the energy change sometimes shows a discrepancy in the forcing terms, as observed in Figure \ref{fig:CV_trace}$b)$ at times 16 to 17.
This can also be seen in the sum at these same times, perhaps showing additional error which occurs when multiple atoms are interacting in a cell.
Despite these minor differences, both momentum and energy conservation validate the forms of stress and heat flux respectively.

\section{Conclusions}
\label{sec:conclusions}
This work has shown the MACE potential can be used for non equilibrium molecular dynamics (NEMD) simulation with local pressure measurements and heat flux obtained using the method of planes (MoP).
These MoP pressure measurements are shown to satisfy the static balance near a liquid-solid interface (a condition where the widely used virial form fails) as well as satisfying the momentum and energy control volume equations.
This conservation demonstrates the MoP pressure is valid arbitrarily far from equilibrium, which means these tools can be directly used in studying fluid dynamics and heat transfer with MACE.
The definition and correct form of pressure is a source of no small controversy in the literature \citep{Shi_et_al(2023)}. 
As a non-unique quantity \citep{Schofield_Henderson}, this has resulted in ambiguity leading to the use of incorrect or inappropriate definitions.
As we move to ML potentials which layer even more ambiguity, including the definition of an effective pairwise force in a graph network, it is useful to try to establish a firmer foundation for the vital quantities like pressure and heat flux.
The form of pressure presented in \eq{CV_pressure} and heat flux in \eq{MOP_energy} have been shown to conserve momentum and energy at every timestep in volumes thin enough to often have only a single atom.
As a result, these validate the MoP forms for use in molecular fluid dynamics systems where the continuum assumption is invalid and far away from thermodynamic equilibrium.
These conservation checks should form an essential part of NEMD validation in the ML age.
The code to run these cases in ASE with MACE is provided along with this work as a template for developing these checks.
The potential for machine learning (ML) to include quantum mechanics (QM) details into molecular fluid simulation are vast.
Many industrial problems studied with NEMD require pressure or heat flux, for example to get slip lengths, heat transfer, solid-liquid interface traction, local visco-elastic effects or liquid-vapour interfacial tension.
This work uses the non-trivial case of a Zirconium dioxide (zirconia) and water interface to demonstrate the MoP method remains applicable in obtaining both the pressure and heat flux.
As the MACE form of potential allows any combination of elements to be modelled and the presented derivation makes no assumptions limiting element type or chemistry \footnote{with the note that long range interactions are currently not included in MACE}, this work shows that NEMD and MACE together have the potential to open a new frontier of molecular fluid dynamics and thermal modelling.

\section*{Acknowledgements}
The author would like to thank Jing Yang \citep{Yang_et_al2021} for sharing her VASP input files which formed the basis for the geometry in this work. The author is grateful to the UK Materials and Molecular Modelling Hub for computational resources, which is partially funded by EPSRC (EP/T022213/1, EP/W032260/1 and EP/P020194/1)).

\section*{Data Availability}
All code used for this project is made available on the \href{https://github.com/edwardsmith999/QM_MoP}{authors GitHub} released under a GPL-3.0 license with persistent URL uploaded to zenodo or similar upon final publication.
This includes numba accelerated code to calculate the MoP in ASE, which could be extended for other systems and ML potentials.



\appendix
\section{Alternative Heat Flux Calculation}
\label{sec:heatflux}

The form which gives the right heat flux in a box \eq{fij_def} is currently prohibitively expensive for anything but very short runs.
One potential approach could use the conservation of energy in each control volume to solve for heat flux.
We can rearrange equation \eq{energy_conservation} written in terms of control volume $I$ with surfaces $I$ on the bottom and $I+1$ on the top,
\begin{align}
\HeatfluxMOP{\!\!\!}_{z^{I+1}}^C  = \HeatfluxMOP{\!\!\!}_{z^{I}}^C + \underbrace{\frac{\partial }{\partial t}  \displaystyle\sum_{i=1}^{N}  e_i  \vartheta_i^I  - \left(\HeatfluxMOP{\!\!\!}_{z^{I+1}}^K - \HeatfluxMOP{\!\!\!}_{z^{I}}^K \right)}_{b_I}
\label{rearrange_heatflux}
\end{align}
From only a single value of $\HeatfluxMOP{\!\!\!}_{z^{I}}^C$, together with the relatively inexpensive kinetic flux and time evolution in a volume, it is possible to use \eq{rearrange_heatflux} to determine all the interaction-based surface fluxes.
Introducing a simple notation $J_I \define \HeatfluxMOP{\!\!\!}_{z^{I}}^C$ so that \eq{rearrange_heatflux} is $J_2 = J_1  + b_1$ for $I=1$, etc, the system of equations can be written as, 
\begin{align}
\begin{bmatrix}
    -1 & 1 & 0 & \cdots & 0 \\
    0 & -1 & 1 & \cdots & 0 \\
    \vdots & & \ddots & \ddots & \vdots \\
    0 & \cdots & 0 & -1 & 1
\end{bmatrix}
\begin{bmatrix}
    J_1 \\
    J_2 \\
    \vdots\\
    J_{N_{bins}}
\end{bmatrix} 
 =     
\begin{bmatrix}
    b_1 \\
    b_2 \\
    \vdots\\
    b_{N_{bins}}
\end{bmatrix}
\label{matrix_J}
\end{align}
In this form, there are $N_{bins}$ equations but $N_{bins}+1$ planes to determine. 
As a result, one additional equation is required, which could be the heat flux on a single plane obtained using only senders and receivers crossing that plane, requiring calls to \texttt{autograd(node\_energy[i], r)} that no longer scale as $N$.
Note this approach would also be applicable to the momentum equation to get configurational pressure; rearranging \eq{MOP_pressure} gives the same system of equation with $J_I \define \StressCV{}_{\!\!{z^I}}^{\!\!C}$ and a different $b_I$.
The difference between the top and bottom values of reconstructed configurational fluxes $\HeatfluxMOP{\!\!\!}_{z^{I+1}}^C - \HeatfluxMOP{\!\!\!}_{z^{I}}^C$ agrees well with the corresponding difference between the top and bottom values calculated from \eq{MOP_energy}.
However, the actual values do not, showing a substantial drift over the domain.
For heat flux, any error in the single measurement are propagated throughout the domain as we calculate the flux value in the next volume from the previous one.
This looks more promising for pressure where the exact agreement to machine precision in each volume ensures agreement, starting from a single bin and using \eq{matrix_J} to get all domain pressures.

This approach would need to be adapted in the presence of external forcing, with SLLOD or Poiseuille flow the work done by the applied force in each cell $F_{i, ext} v_i$ should be added to the balance equation in $b_I$. Similarly, wall tethering forces would also need to be added when studying wall driven Couette flow. Thermostatting or Barostatting would also need to be included in the energy balance. However, these constraints are often applied to just a part of the NEMD simulation cell, so this approach could be used on a subregion of the domain without external forces (for example only the fluid), with the reference cell chosen as the edge of this subregion.

\bibliography{./ref}

@BOOK{Goldstein,
	author	  =  {H. Goldstein and C. Poole and J. Safko},
	title	  =  {Classical Mechanics},
	publisher =  {Addison Wesley},
	year	  =  {2002},
	edition   =  {3rd}
}

@BOOK{Evans_Morris,
	author	  =  {D. J. Evans and G. P. Morris},
	title	  =  {Statistical Mechanics of Non-Equilibrium Liquids},
	publisher =  {Australian National University Press},
	year	  =  {2007},
	edition   =  {2nd}
}

@article{Irving_kirkwood,
	author	  =  {J. H. Irving and J. G. Kirkwood},
	title	  =  {The Statistical Mechanics Theory of Transport Processes. IV. The Equations of Hydrodynamics},
	journal   =  {J. Chemical. Phys.},
	year	  =  {1950},
	volume    =  {18 - 6},
	pages     =  {817-829}
}

@article{Hardy,
	author	  =  {R. J. Hardy},
	title	  =  {Formulas for determining local properties in molecular dynamics simulations: Shock waves},
	journal   =  {J. Chem. Phys},
	year	  =  {1982},
	volume    =  {76(1)},
	pages     =  {622-628}
}

@article{Todd_et_al_95,
	author	  =  {B.D. Todd and D.J. Evans and P.J. Daivis},
	title	  =  {Pressure Tensor for inhomogenous fluids},
	journal   =  {Physical Review E},
	year	  =  {1995},
	volume    =  {52(2)},
	pages     =  {1627-1638}
}

@book{todd_daivis_2017,
	Author = {Todd, Billy D. and Daivis, Peter J.},
	Place = {Cambridge},
	Publisher = {Cambridge University Press},
	Title = {Nonequilibrium Molecular Dynamics: Theory, Algorithms and Applications},
	Year = {2017}}

@article{Schofield_Henderson,
	Author = {Schofield, P. and Henderson, J. R.},
	Journal = {Proc. R. Soc. Lond. A},
	Pages = {231},
	Title = {{Statistical Mechanics of Inhomogeneous Fluids}},
	Volume = {379},
	Year = {1982}}

@article{Cormier_et_al,
	Author = {Cormier, J. and Rickman, J. M. and Delph, T. J.},
	Journal = {J. Appl. Phys.},
	Pages = {99},
	Title = {{Stress calculation in atomistic simulations of perfect and imperfect solids}},
	Volume = {89},
	Year = {2001}}

@article{Han_Lee,
	Author = {Han, M. and Lee, J. S.},
	Journal = {Phys. Rev. E},
	Pages = {061205},
	Title = {{Method for calculating the heat and momentum fluxes of inhomogeneous fluids}},
	Volume = {70},
	Year = {2004}}

@article{Smith_et_al12,
	Author = {Smith, E. R. and Heyes, D. M. and Dini, D. and Zaki, T. A.},
	Journal = {Phys. Rev. E.},
	Pages = {056705},
	Title = {{Control-volume representation of molecular dynamics}},
	Volume = {85},
	Year = {2012}}

@MANUAL{LAMMPS,
	title	     =  {LAMMPS Users Manual - Large-scale Atomic/Molecular Massively Parallel Simulator},
	author	     =  {S. Plimpton and P. Crozier and A. Thompson},
	organisation =  {Sandia Corporation},
	address      =  {http://lammps.sandia.gov},
	edition	     =  {7 Jul 2009 version},
	year	     =  {2003}
}

@article{Monaghan1997,
  title = {Microscopic study of steady convective flow in periodic systems},
  author = {Monaghan, David R. J. and Morriss, Gary P.},
  journal = {Physical Review E},
  volume = {56},
  number = {1},
  pages = {476}, 
  year = {1997},
  month = jul,
  publisher = {American Physical Society},
}

@article{Eyert2023,
  author    = {Volker Eyert and Jonathan Wormald and William A. Curtin and Erich Wimmer},
  title     = {Machine-learned interatomic potentials: Recent developments and prospective applications},
  journal   = {Journal of Materials Research},
  volume    = {38},
  pages     = {5079--5094},
  year      = {2023},
  doi       = {10.1557/s43578-023-01239-8}
}

@article{Noll1955,
  author  = {Noll, W.},
  title   = {Die Herleitung der Grundgleichungen der Thermomechanik der Kontinua aus der statistischen Mechanik},
  journal = {Journal of Rational Mechanics and Analysis},
  volume  = {4},
  pages   = {627--646},
  year    = {1955},
}

@article{Parker,
	Author = {Parker, E. N.},
	Journal = {Phys. Rev.},
	Pages = {1686},
	Title = {{Tensor Virial Equations}},
	Volume = {96},
	Year = {1954}}

@article{Kirkwood_Buff,
	Author = {Kirkwood, John G. and Buff, Frank P.},
	Journal = {The Journal of Chemical Physics},
	Number = {3},
	Pages = {338-343},
	Title = {The Statistical Mechanical Theory of Surface Tension},
	Volume = {17},
	Year = {1949}}

@article{langer2023stress,
  author       = {Langer, Marcel F. and Frank, J. Thorben and Knoop, Florian},
  title        = {Stress and heat flux via automatic differentiation},
  journal      = {arXiv preprint arXiv:2305.01401},
  year         = {2023},
  url          = {https://doi.org/10.48550/arXiv.2305.01401},
  note         = {arXiv:2305.01401}
}

@article{Langer2023heatflux,
  title = {Heat flux for semilocal machine-learning potentials},
  author = {Langer, Marcel F. and Knoop, Florian and Carbogno, Christian and Scheffler, Matthias and Rupp, Matthias},
  journal = {Phys. Rev. B},
  volume = {108},
  issue = {10},
  pages = {L100302},
  numpages = {7},
  year = {2023},
  month = {Sep},
  publisher = {American Physical Society},
  doi = {10.1103/PhysRevB.108.L100302},
  url = {https://link.aps.org/doi/10.1103/PhysRevB.108.L100302}
}

@article{deng_2023_chgnet,
title={CHGNet as a pretrained universal neural network potential for charge-informed atomistic modelling},
DOI={10.1038/s42256-023-00716-3},
journal={Nature Machine Intelligence},
author={Deng, Bowen and Zhong, Peichen and Jun, KyuJung and Riebesell, Janosh and Han, Kevin and Bartel, Christopher J. and Ceder, Gerbrand},
year={2023},
pages={1-11}
}

@misc{barroso_omat24,
      title={Open Materials 2024 (OMat24) Inorganic Materials Dataset and Models}, 
      author={Luis Barroso-Luque and Muhammed Shuaibi and Xiang Fu and Brandon M. Wood and Misko Dzamba and Meng Gao and Ammar Rizvi and C. Lawrence Zitnick and Zachary W. Ulissi},
      year={2024},
      eprint={2410.12771},
      archivePrefix={arXiv},
      primaryClass={cond-mat.mtrl-sci},
      url={https://arxiv.org/abs/2410.12771}, 
}

@article{batatia2022mace,
  author       = {Batatia, Ilyes and Kovacs, Dániel P. and Ortner, Christoph and Cs{\'a}nyi, G{\'a}bor and Deringer, Volker L.},
  title        = {MACE: Higher order equivariant message passing neural networks for fast and accurate force fields},
  journal      = {arXiv preprint arXiv:2206.07697},
  year         = {2022},
  url          = {https://arxiv.org/abs/2206.07697},
  note         = {arXiv:2206.07697}
}

@article{batatia2023foundation,
      title={A foundation model for atomistic materials chemistry},
      author={Ilyes Batatia and Philipp Benner and Yuan Chiang and Alin M. Elena and Dávid P. Kovács and Janosh Riebesell and Xavier R. Advincula and Mark Asta and William J. Baldwin and Noam Bernstein and Arghya Bhowmik and Samuel M. Blau and Vlad Cărare and James P. Darby and Sandip De and Flaviano Della Pia and Volker L. Deringer and Rokas Elijošius and Zakariya El-Machachi and Edvin Fako and Andrea C. Ferrari and Annalena Genreith-Schriever and Janine George and Rhys E. A. Goodall and Clare P. Grey and Shuang Han and Will Handley and Hendrik H. Heenen and Kersti Hermansson and Christian Holm and Jad Jaafar and Stephan Hofmann and Konstantin S. Jakob and Hyunwook Jung and Venkat Kapil and Aaron D. Kaplan and Nima Karimitari and Namu Kroupa and Jolla Kullgren and Matthew C. Kuner and Domantas Kuryla and Guoda Liepuoniute and Johannes T. Margraf and Ioan-Bogdan Magdău and Angelos Michaelides and J. Harry Moore and Aakash A. Naik and Samuel P. Niblett and Sam Walton Norwood and Niamh O'Neill and Christoph Ortner and Kristin A. Persson and Karsten Reuter and Andrew S. Rosen and Lars L. Schaaf and Christoph Schran and Eric Sivonxay and Tamás K. Stenczel and Viktor Svahn and Christopher Sutton and Cas van der Oord and Eszter Varga-Umbrich and Tejs Vegge and Martin Vondrák and Yangshuai Wang and William C. Witt and Fabian Zills and Gábor Csányi},
      year={2023},
	  journal   = {arXiv preprint arXiv},
      eprint={2401.00096},
      archivePrefix={arXiv},
      primaryClass={physics.chem-ph}
}

@article{RDrautz_2004,
	doi = {10.1088/0953-8984/16/23/005},
	year = {2004},
	month = {may},
	publisher = {},
    journal   = {J. Phys. Condens. Matter},
	volume = {16},
	number = {23},
	pages = {3843},
	author = {R Drautz and M Fähnle and J M Sanchez},
	title = {General relations between many-body potentials and cluster expansions in multicomponent
	systems},
}

@article{kaplan2025foundational,
  author    = {Kaplan, Alexander D. and Liu, Ruoxi and Qi, Junqi and Ko, Tae Wook and Deng, Bowen and Riebesell, Julian and Ceder, Gerbrand and Persson, Kristin A. and Ong, Shyue Ping},
  title     = {A Foundational Potential Energy Surface Dataset for Materials},
  journal   = {arXiv preprint arXiv:2503.04070},
  year      = {2025},
  doi       = {10.48550/arXiv.2503.04070},
  url       = {https://arxiv.org/abs/2503.04070}
}

@article{Admal_Tadmore_2010,
	Archiveprefix = {arXiv},
	Arxivid = {1008.4819},
	Author = {Admal, Nikhil Chandra and Tadmor, E. B.},
	Doi = {10.1007/s10659-010-9249-6},
	Eprint = {1008.4819},
	Journal = {Journal of Elasticity},
	Number = {1-2},
	Pages = {63--143},
	Title = {{A unified interpretation of stress in molecular systems}},
	Volume = {100},
	Year = {2010}
}

@article{fan2015force,
  author    = {Fan, Zheyong and Pereira, Luiz Felipe C. and Wang, Hui‑Qiong and Zheng, Jin‑Cheng and Donadio, Davide and Harju, Ari},
  title     = {Force and heat current formulas for many-body potentials in molecular dynamics simulation with applications to thermal conductivity calculations},
  journal   = {Physical Review B},
  volume    = {92},
  number    = {9},
  pages     = {094301},
  year      = {2015},
  doi       = {10.1103/PhysRevB.92.094301},
  url       = {https://doi.org/10.1103/PhysRevB.92.094301},
}

@article{Admal_Tadmor_2011,
    author = {Admal, Nikhil Chandra and Tadmor, E. B.},
    title = {Stress and heat flux for arbitrary multibody potentials: A unified framework},
    journal = {The Journal of Chemical Physics},
    volume = {134},
    number = {18},
    pages = {184106},
    year = {2011},
    month = {05},
    issn = {0021-9606},
    doi = {10.1063/1.3582905},
    url = {https://doi.org/10.1063/1.3582905},
    eprint = {https://pubs.aip.org/aip/jcp/article-pdf/doi/10.1063/1.3582905/14061374/184106\_1\_online.pdf},
}

@article{Focassio_et_al2025,
    author = {Focassio, Bruno and M. Freitas, Luis Paulo and Schleder, Gabriel R.},
    title = {Performance Assessment of Universal Machine Learning Interatomic Potentials: Challenges and Directions for Materials’ Surfaces},
    journal = {ACS Applied Materials \& Interfaces},
    volume = {17},
    number = {9},
    pages = {13111-13121},
    year = {2025},
    doi = {10.1021/acsami.4c03815},
    note ={PMID: 38990833},

}

@article{heyes2011equivalence,
  title={The equivalence between volume averaging and method of planes definitions of the pressure tensor at a plane},
  author={Heyes, DM and Smith, ER and Dini, D and Zaki, TA},
  journal={The Journal of chemical physics},
  volume={135},
  number={2},
  pages={024512},
  year={2011},
  publisher={AIP},
  doi={10.1063/1.3605692}
}

@article{Smith_et_al19,
    author = {Smith, E. R.  and Daivis, P. J.  and Todd, B. D. },
    title = {Measuring heat flux beyond Fourier’s law},
    journal = {The Journal of Chemical Physics},
    volume = {150},
    number = {6},
    pages = {064103},
    year = {2019},
    doi = {10.1063/1.5079993},
}

@article{Christensen_et_al98,
  title = {First-principles study of the surfaces of zirconia},
  author = {Christensen, A. and Carter, Emily A.},
  journal = {Phys. Rev. B},
  volume = {58},
  issue = {12},
  pages = {8050--8064},
  numpages = {0},
  year = {1998},
  month = {Sep},
  publisher = {American Physical Society},
  doi = {10.1103/PhysRevB.58.8050},
  url = {https://link.aps.org/doi/10.1103/PhysRevB.58.8050}
}

@article{D3_correction,
    author = {Grimme, Stefan and Antony, Jens and Ehrlich, Stephan and Krieg, Helge},
    title = {A consistent and accurate ab initio parametrization of density functional dispersion correction (DFT-D) for the 94 elements H-Pu},
    journal = {The Journal of Chemical Physics},
    volume = {132},
    number = {15},
    pages = {154104},
    year = {2010},
    month = {04},
    issn = {0021-9606},
    doi = {10.1063/1.3382344},
    url = {https://doi.org/10.1063/1.3382344},
}

@article{Lee_et_al94,
    author = {Lee, Chengteh and Chen, Han and Fitzgerald, George},
    title = {Structures of the water hexamer using density functional methods},
    journal = {The Journal of Chemical Physics},
    volume = {101},
    number = {5},
    pages = {4472-4473},
    year = {1994},
    month = {09},
    issn = {0021-9606},
    doi = {10.1063/1.467434},
    url = {https://doi.org/10.1063/1.467434},
}

@article{batatia2025,
  title={Cross Learning between Electronic Structure Theories for Unifying Molecular, Surface, and Inorganic Crystal Foundation Force Fields},
  author={Batatia, Ilyes and Lin, Chen and Hart, Joseph and Kasoar, Elliott and Elena, Alin M and Norwood, Sam Walton and Wolf, Thomas and Cs{\u{A}}{\k{A}}nyi, G{\u{A}}{\k{A}}bor},
  journal={arXiv preprint arXiv:2510.25380},
  year={2025}
}

@article{batzner2022e3gnn,
  author    = {Batzner, Simon and Musaelian, Albert and Sun, Lixin and others},
  title     = {E(3)-equivariant graph neural networks for data-efficient and accurate interatomic potentials},
  journal   = {Nature Communications},
  volume    = {13},
  pages     = {2453},
  year      = {2022},
  doi       = {10.1038/s41467-022-29939-5},
}

@article{Kovacs_et_al2023,
    author = {Kov{\'a}cs, D{\'a}vid P{\'e}ter and Batatia, Ilyes and Arany, Eszter S{\'a}ra and Cs{\'a}nyi, Gábor},
    title = {Evaluation of the MACE force field architecture: From medicinal chemistry to materials science},
    journal = {The Journal of Chemical Physics},
    volume = {159},
    number = {4},
    pages = {044118},
    year = {2023},
    month = {07},
    issn = {0021-9606},
    doi = {10.1063/5.0155322},
}

@misc{thomas2018tensorfieldnetworks,
    title={Tensor field networks: Rotation- and translation-equivariant neural networks for 3D point clouds}, 
    author={Nathaniel Thomas and Tess Smidt and Steven Kearnes and Lusann Yang and Li Li and Kai Kohlhoff and Patrick Riley},
    year={2018},
    eprint={1802.08219},
    archivePrefix={arXiv},
    primaryClass={cs.LG},
    url={https://arxiv.org/abs/1802.08219}
}

@misc{weiler20183dsteerablecnns,
    title={3D Steerable CNNs: Learning Rotationally Equivariant Features in Volumetric Data}, 
    author={Maurice Weiler and Mario Geiger and Max Welling and Wouter Boomsma and Taco Cohen},
    year={2018},
    eprint={1807.02547},
    archivePrefix={arXiv},
    primaryClass={cs.LG},
    url={https://arxiv.org/abs/1807.02547}
}

@misc{kondor2018clebschgordannets,
    title={Clebsch-Gordan Nets: a Fully Fourier Space Spherical Convolutional Neural Network}, 
    author={Risi Kondor and Zhen Lin and Shubhendu Trivedi},
    year={2018},
    eprint={1806.09231},
    archivePrefix={arXiv},
    primaryClass={stat.ML},
    url={https://arxiv.org/abs/1806.09231}
}

@misc{e3nn_software,
    author = {Mario Geiger and
              Tess Smidt and
              Alby M. and
              Benjamin Kurt Miller and
              Wouter Boomsma and
              Bradley Dice and
              Kostiantyn Lapchevskyi and
              Maurice Weiler and
              Michał Tyszkiewicz and
              Simon Batzner and
              Dylan Madisetti and
              Martin Uhrin and
              Jes Frellsen and
              Nuri Jung and
              Sophia Sanborn and
              Mingjian Wen and
              Josh Rackers and
              Marcel Rød and
              Michael Bailey},
    title = {Euclidean neural networks: e3nn},
    month = apr,
    year = 2022,
    publisher = {Zenodo},
    version = {0.5.0},
    doi = {10.5281/zenodo.6459381},
    url = {https://doi.org/10.5281/zenodo.6459381}
}

@article{Senftle2016,
  author    = {Senftle, Thomas P. and Hong, Sungwook and Islam, Md Mahbubul and Kylasa, Sudhir B. and Zheng, Yuanxia and Shin, Yun Kyung and Junkermeier, Chad and Engel-Herbert, Roman and Janik, Michael J. and Aktulga, Hasan Metin and Verstraelen, Toon and Grama, Ananth and van Duin, Adri C. T.},
  title     = {The ReaxFF reactive force-field: development, applications and future directions},
  journal   = {npj Computational Materials},
  volume    = {2},
  number    = {1},
  pages     = {15011},
  year      = {2016},
  doi       = {10.1038/npjcompumats.2015.11},
  url       = {https://doi.org/10.1038/npjcompumats.2015.11},
  issn      = {2057-3960}
}

@article{Green,
  author    = {Green, Melville S.},
  title     = {Markoff random processes and the statistical mechanics of time-dependent phenomena. II. Irreversible processes in fluids},
  journal   = {Journal of Chemical Physics},
  volume    = {22},
  number    = {3},
  pages     = {398--413},
  year      = {1954},
  doi       = {10.1063/1.1740082}
}

@article{Kubo,
  author    = {Kubo, Ryogo},
  title     = {Statistical-mechanical theory of irreversible processes. I. General theory and simple applications to magnetic and conduction problems},
  journal   = {Journal of the Physical Society of Japan},
  volume    = {12},
  number    = {6},
  pages     = {570--586},
  year      = {1957},
  doi       = {10.1143/JPSJ.12.570}
}

@article{Ewen2017,
  author    = {Ewen, J. P. and Gattinoni, C. and Zhang, J. and Heyes, D. M. and Spikes, H. A. and Dini, D.},
  title     = {On the effect of confined fluid molecular structure on nonequilibrium phase behaviour and friction},
  journal   = {Physical Chemistry Chemical Physics},
  volume    = {19},
  number    = {27},
  pages     = {17883--17894},
  year      = {2017},
  doi       = {10.1039/C7CP02638A}
}

@article{Kannam2012,
  author    = {Kannam, S. Kumar and Todd, B. D. and Hansen, J. S. and Daivis, P. J.},
  title     = {Slip length of water on graphene: Limitations of non-equilibrium molecular dynamics simulations},
  journal   = {Journal of Chemical Physics},
  volume    = {136},
  number    = {2},
  pages     = {024705},
  year      = {2012},
  doi       = {10.1063/1.3675904}
}

@article{Williams2022,
  author    = {Williams, David and Wei, Zhenyu and Shaharudin, Mohd R. b. and Carbone, Piero},
  title     = {A molecular simulation study into the stability of hydrated graphene nanochannels used in nanofluidics devices},
  journal   = {Nanoscale},
  volume    = {14},
  number    = {9},
  pages     = {3467--3479},
  year      = {2022},
  doi       = {10.1039/D1NR07045E}
}

@article{Thompson1989,
  author    = {Thompson, A. P. and Robbins, Mark O.},
  title     = {Simulations of contact-line motion: Slip and the dynamic contact angle},
  journal   = {Physical Review Letters},
  volume    = {63},
  number    = {7},
  pages     = {766--769},
  year      = {1989},
  doi       = {10.1103/PhysRevLett.63.766}
}

@article{Qian2005,
  author    = {Qian, Tiezheng and Wang, Xiao-Ping and Sheng, Ping},
  title     = {Molecular hydrodynamics of the moving contact line in two-phase immiscible flows},
  journal   = {Communications in Computational Physics},
  volume    = {1},
  pages     = {1--52},
  year      = {2005},
  eprint    = {cond-mat/0510403},
  archivePrefix = {arXiv}
}

@article{OConnell1995,
  author    = {O'Connell, S. T. and Thompson, P. A.},
  title     = {Molecular dynamics-continuum hybrid computations: A tool for studying complex fluid flows},
  journal   = {Physical Review E},
  volume    = {52},
  number    = {6},
  pages     = {R5792--R5795},
  year      = {1995},
  doi       = {10.1103/PhysRevE.52.R5792}
}

@article{Mohamed2010,
  author    = {Mohamed, M. and Mohamad, A. A.},
  title     = {A review of the development of hybrid atomistic-continuum methods for dense fluids},
  journal   = {Microfluidics and Nanofluidics},
  volume    = {8},
  number    = {3},
  pages     = {283--302},
  year      = {2010},
  doi       = {10.1007/s10404-009-0521-2}
}

@article{Flekkoy2000,
  author    = {Flekk{\o}y, Eirik G. and Wagner, Gunnar and Feder, Jens},
  title     = {Hybrid model for combined particle and continuum dynamics},
  journal   = {Europhysics Letters},
  volume    = {52},
  number    = {3},
  pages     = {271--276},
  year      = {2000},
  doi       = {10.1209/epl/i2000-00417-8}
}

@article{Davydov2014,
  author    = {Davydov, Denis and Pelteret, Jean-Paul and Steinmann, Paul},
  title     = {Comparison of several staggered atomistic-to-continuum concurrent coupling strategies},
  journal   = {Computer Methods in Applied Mechanics and Engineering},
  volume    = {277},
  pages     = {260--280},
  year      = {2014},
  doi       = {10.1016/j.cma.2014.04.010}
}

@article{Hammonds_Heyes2021,
    author = {Hammonds, K. D. and Heyes, D. M.},
    title = {Shadow Hamiltonian in classical NVE molecular dynamics simulations involving Coulomb interactions},
    journal = {The Journal of Chemical Physics},
    volume = {154},
    number = {17},
    pages = {174102},
    year = {2021},
    month = {05},
    doi = {10.1063/5.0048194},
}

@article{Yang_et_al2021,
	author = {Yang, Jing and Youssef, Mostafa and Yildiz, Bilge},
	title = {Structure, Kinetics, and Thermodynamics of Water and Its Ions at the Interface with Monoclinic ZrO2 Resolved via Ab Initio Molecular Dynamics},
	journal = {The Journal of Physical Chemistry C},
	volume = {125},
	number = {28},
	pages = {15233-15242},
	year = {2021},
	doi = {10.1021/acs.jpcc.1c02064},
}

@inproceedings{Numba,
author = {Lam, Siu Kwan and Pitrou, Antoine and Seibert, Stanley},
title = {Numba: a LLVM-based Python JIT compiler},
year = {2015},
isbn = {9781450340052},
publisher = {Association for Computing Machinery},
address = {New York, NY, USA},
url = {https://doi.org/10.1145/2833157.2833162},
doi = {10.1145/2833157.2833162},
booktitle = {Proceedings of the Second Workshop on the LLVM Compiler Infrastructure in HPC},
articleno = {7},
numpages = {6},
location = {Austin, Texas},
series = {LLVM '15}
}

@misc{cuEquivariance,
    url = {https://github.com/NVIDIA/cuEquivariance},
    author = {NVIDIA},
    title = {cuEquivariance},
    publisher = {website},
    year = {2025},
    copyright = {Apache 2.0}
}

@article{ase-paper,
  author={Ask Hjorth Larsen and Jens Jørgen Mortensen and Jakob Blomqvist and Ivano E Castelli and Rune Christensen and Marcin
Dułak and Jesper Friis and Michael N Groves and Bjørk Hammer and Cory Hargus and Eric D Hermes and Paul C Jennings and Peter
Bjerre Jensen and James Kermode and John R Kitchin and Esben Leonhard Kolsbjerg and Joseph Kubal and Kristen
Kaasbjerg and Steen Lysgaard and Jón Bergmann Maronsson and Tristan Maxson and Thomas Olsen and Lars Pastewka and Andrew
Peterson and Carsten Rostgaard and Jakob Schiøtz and Ole Schütt and Mikkel Strange and Kristian S Thygesen and Tejs
Vegge and Lasse Vilhelmsen and Michael Walter and Zhenhua Zeng and Karsten W Jacobsen},
  title={The atomic simulation environment—a Python library for working with atoms},
  journal={Journal of Physics: Condensed Matter},
  volume={29},
  number={27},
  pages={273002},
  url={http://stacks.iop.org/0953-8984/29/i=27/a=273002},
  year={2017}
}

@article{Heinz_et_al05,
  title = {Calculation of local pressure tensors in systems with many-body interactions},
  author = {Heinz, Hendrik and Paul, Wolfgang and Binder, Kurt},
  journal = {Phys. Rev. E},
  volume = {72},
  issue = {6},
  pages = {066704},
  numpages = {10},
  year = {2005},
  month = {Dec},
  publisher = {American Physical Society},
  doi = {10.1103/PhysRevE.72.066704},
  url = {https://link.aps.org/doi/10.1103/PhysRevE.72.066704}
}

@article{Heinz2007,
    author = {H. Heinz},
    title = {Calculation of local and average pressure tensors in molecular simulations},
    journal = {Molecular Simulation},
    volume = {33},
    number = {9-10},
    pages = {747--758},
    year = {2007},
    publisher = {Taylor \& Francis},
    doi = {10.1080/08927020701308828},
}

@phdthesis{admal2014results,
  title={Results on the interaction between atomistic and continuum models},
  author={Admal, Nikhil Chandra},
  year={2014},
  school={University of Minnesota}
}

@article{Langer_2024,
doi = {10.1088/2632-2153/ad86a0},
url = {https://doi.org/10.1088/2632-2153/ad86a0},
year = {2024},
month = {oct},
publisher = {IOP Publishing},
volume = {5},
number = {4},
pages = {04LT01},
author = {Langer, Marcel F and Pozdnyakov, Sergey N and Ceriotti, Michele},
title = {Probing the effects of broken symmetries in machine learning},
journal = {Machine Learning: Science and Technology},
}

@article{Brandstetter2021GeometricAP,
  title={Geometric and Physical Quantities improve E(3) Equivariant Message Passing},
  author={Johannes Brandstetter and Rob D. Hesselink and Elise van der Pol and Erik J. Bekkers and Max Welling},
  journal={ArXiv},
  year={2021},
  volume={abs/2110.02905},
  url={https://api.semanticscholar.org/CorpusID:238407870}
}

\end{document}